\pdfoutput=1
\documentclass[12pt,a4paper]{article}

\usepackage{ifthen} 
\newboolean{pdflatex}
\setboolean{pdflatex}{true} 

\newboolean{articletitles}
\setboolean{articletitles}{true} 

\newboolean{uprightparticles}
\setboolean{uprightparticles}{false} 

\def\paperauthors{LHCb collaboration} 
\def\paperasciititle{Search for Bc+->pi+mu+mu- decays and measurement of the branching fraction ratio BF(Bc+->psi(2S) pi+)/BF(Bc+->J/psi pi+)} 
\def\papertitle{Search for \decay{\Bc}{\pip\mumu} decays and measurement of the\\ branching fraction ratio $\BF(\decay{\Bc}{\psitwos\pip})/\BF(\decay{\Bc}{\jpsi\pip})$} 
\def\paperkeywords{{High Energy Physics}, {LHCb}} 
\def\papercopyright{\the\year\ CERN for the benefit of the LHCb collaboration} 
\def\paperlicence{CC BY 4.0 licence}
\def\paperlicenceurl{https://creativecommons.org/licenses/by/4.0/}

\usepackage{comment}
\usepackage{booktabs}

\usepackage[top=1in, bottom=1.25in, left=1in, right=1in]{geometry}

\usepackage{microtype}
\usepackage{lineno}  
\usepackage{xspace} 
\usepackage{caption} 

\usepackage{graphicx}  
\usepackage{color}
\usepackage{colortbl}
\graphicspath{{./figs/}} 

\usepackage{amsmath} 
\usepackage{amssymb}
\usepackage{amsfonts}
\usepackage{upgreek} 

\newcommand*\patchAmsMathEnvironmentForLineno[1]{%
\expandafter\let\csname old#1\expandafter\endcsname\csname #1\endcsname
\expandafter\let\csname oldend#1\expandafter\endcsname\csname
end#1\endcsname
 \renewenvironment{#1}%
   {\linenomath\csname old#1\endcsname}%
   {\csname oldend#1\endcsname\endlinenomath}%
}
\newcommand*\patchBothAmsMathEnvironmentsForLineno[1]{%
  \patchAmsMathEnvironmentForLineno{#1}%
  \patchAmsMathEnvironmentForLineno{#1*}%
}
\AtBeginDocument{%
\patchBothAmsMathEnvironmentsForLineno{equation}%
\patchBothAmsMathEnvironmentsForLineno{align}%
\patchBothAmsMathEnvironmentsForLineno{flalign}%
\patchBothAmsMathEnvironmentsForLineno{alignat}%
\patchBothAmsMathEnvironmentsForLineno{gather}%
\patchBothAmsMathEnvironmentsForLineno{multline}%
\patchBothAmsMathEnvironmentsForLineno{eqnarray}%
}

\usepackage{hyperxmp}

\usepackage[pdftex,
            pdfauthor={\paperauthors},
            pdftitle={\paperasciititle},
            pdfkeywords={\paperkeywords},
            pdfcopyright={Copyright (C) \papercopyright},
            pdflicenseurl={\paperlicenceurl}]{hyperref}

\usepackage[bottom,flushmargin,hang,multiple]{footmisc}

\usepackage[all]{hypcap} 

\usepackage{xspace} 
\usepackage{upgreek}

\def\lhcb   {\mbox{LHCb}\xspace}

\def\MagUp {\mbox{\em Mag\kern -0.05em Up}\xspace}

\ifthenelse{\boolean{uprightparticles}}%
{

 \def\Pmu         {\ensuremath{\upmu}\xspace}

 \def\Ppi         {\ensuremath{\uppi}\xspace}                 
                  
 \def\Prho        {\ensuremath{\uprho}\xspace}

 \def\Ppsi        {\ensuremath{\uppsi}\xspace}

 \def\PDelta      {\ensuremath{\Delta}\xspace}                 
 \def\PXi         {\ensuremath{\Xi}\xspace}                 
 \def\PLambda     {\ensuremath{\Lambda}\xspace}                 
 \def\PSigma      {\ensuremath{\Sigma}\xspace}                 
 \def\POmega      {\ensuremath{\Omega}\xspace}                 
 \def\PUpsilon    {\ensuremath{\Upsilon}\xspace}
 \let\oldPi\Pi
 \def\PPi         {\ensuremath{\oldPi}\xspace}

 \def\PB      {\ensuremath{\mathrm{B}}\xspace}                 
                  
 \def\PD      {\ensuremath{\mathrm{D}}\xspace}

 \def\PJ      {\ensuremath{\mathrm{J}}\xspace}                 
 \def\PK      {\ensuremath{\mathrm{K}}\xspace}

 \def\PW      {\ensuremath{\mathrm{W}}\xspace}

 \def\PZ      {\ensuremath{\mathrm{Z}}\xspace}                 
                  
 \def\Pb      {\ensuremath{\mathrm{b}}\xspace}                 
 \def\Pc      {\ensuremath{\mathrm{c}}\xspace}                 
 \def\Pd      {\ensuremath{\mathrm{d}}\xspace}                 
 \def\Pe      {\ensuremath{\mathrm{e}}\xspace}

 \def\Pi      {\ensuremath{\mathrm{i}}\xspace}

 \def\Ps      {\ensuremath{\mathrm{s}}\xspace}                 
                  
 \def\Pu      {\ensuremath{\mathrm{u}}\xspace}

 \def\thebaroffset{0.0em}
}
{

 \def\Pmu         {\ensuremath{\mu}\xspace}

 \def\Ppi         {\ensuremath{\pi}\xspace}                 
                  
 \def\Prho        {\ensuremath{\rho}\xspace}

 \def\Ppsi        {\ensuremath{\psi}\xspace}                 
                  
 \mathchardef\PDelta="7101
 \mathchardef\PXi="7104
 \mathchardef\PLambda="7103
 \mathchardef\PSigma="7106
 \mathchardef\POmega="710A
 \mathchardef\PUpsilon="7107
 \mathchardef\PPi="7105
                  
 \def\PB      {\ensuremath{B}\xspace}                 
                  
 \def\PD      {\ensuremath{D}\xspace}

 \def\PJ      {\ensuremath{J}\xspace}                 
 \def\PK      {\ensuremath{K}\xspace}

 \def\PW      {\ensuremath{W}\xspace}

 \def\PZ      {\ensuremath{Z}\xspace}                 
                  
 \def\Pb      {\ensuremath{b}\xspace}                 
 \def\Pc      {\ensuremath{c}\xspace}                 
 \def\Pd      {\ensuremath{d}\xspace}                 
 \def\Pe      {\ensuremath{e}\xspace}

 \def\Pi      {\ensuremath{i}\xspace}

 \def\Ps      {\ensuremath{s}\xspace}                 
                  
 \def\Pu      {\ensuremath{u}\xspace}

 \def\thebaroffset{0.18em}
}
\newcommand{\offsetoverline}[2][\thebaroffset]{\kern #1\overline{\kern -#1 #2}}%

\makeatletter
\ifcase \@ptsize \relax
  \newcommand{\miniscule}{\@setfontsize\miniscule{4}{5}}
\or
  \newcommand{\miniscule}{\@setfontsize\miniscule{5}{6}}
\or
  \newcommand{\miniscule}{\@setfontsize\miniscule{5}{6}}
\fi
\makeatother

\DeclareRobustCommand{\optbar}[1]{\shortstack{{\miniscule (\rule[.5ex]{1.25em}{.18mm})}
  \\ [-.7ex] $#1$}}

\def\en         {{\ensuremath{\Pe^-}}\xspace}   
\def\ep         {{\ensuremath{\Pe^+}}\xspace}

\def\epem       {{\ensuremath{\Pe^+\Pe^-}}\xspace}

\def\mup        {{\ensuremath{\Pmu^+}}\xspace}

\def\mumu       {{\ensuremath{\Pmu^+\Pmu^-}}\xspace}

\def\W      {{\ensuremath{\PW}}\xspace}

\def\Z      {{\ensuremath{\PZ}}\xspace}

\def\uquark    {{\ensuremath{\Pu}}\xspace}

\def\dquark    {{\ensuremath{\Pd}}\xspace}
\def\dquarkbar {{\ensuremath{\overline \dquark}}\xspace}

\def\squark    {{\ensuremath{\Ps}}\xspace}

\def\cquark    {{\ensuremath{\Pc}}\xspace}

\def\bquark    {{\ensuremath{\Pb}}\xspace}
\def\bquarkbar {{\ensuremath{\overline \bquark}}\xspace}

\def\pion   {{\ensuremath{\Ppi}}\xspace}
\def\piz    {{\ensuremath{\pion^0}}\xspace}
\def\pip    {{\ensuremath{\pion^+}}\xspace}
\def\pim    {{\ensuremath{\pion^-}}\xspace}

\def\rhomeson {{\ensuremath{\Prho}}\xspace}

\def\rhop     {{\ensuremath{\rhomeson^+}}\xspace}

\def\kaon    {{\ensuremath{\PK}}\xspace}

\def\KorKbar {\kern \thebaroffset\optbar{\kern -\thebaroffset \PK}{}\xspace}

\def\Kp      {{\ensuremath{\kaon^+}}\xspace}

\def\D       {{\ensuremath{\PD}}\xspace}

\def\DorDbar {\kern \thebaroffset\optbar{\kern -\thebaroffset \PD}\xspace}

\def\Dp      {{\ensuremath{\D^+}}\xspace}
\def\Dm      {{\ensuremath{\D^-}}\xspace}

\def\DpDm    {\ensuremath{\Dp {\kern -0.16em \Dm}}\xspace}

\def\Dsp     {{\ensuremath{\D^+_\squark}}\xspace}

\def\B       {{\ensuremath{\PB}}\xspace}

\def\BorBbar {\kern \thebaroffset\optbar{\kern -\thebaroffset \PB}\xspace}

\def\Bd      {{\ensuremath{\B^0}}\xspace}

\def\BdorBdbar {\kern \thebaroffset\optbar{\kern -\thebaroffset \Bd}\xspace}
\def\Bu      {{\ensuremath{\B^+}}\xspace}

\def\Bp      {{\ensuremath{\Bu}}\xspace}

\def\Bs      {{\ensuremath{\B^0_\squark}}\xspace}

\def\BsorBsbar {\kern \thebaroffset\optbar{\kern -\thebaroffset \Bs}\xspace}
\def\Bc      {{\ensuremath{\B_\cquark^+}}\xspace}

\def\jpsi     {{\ensuremath{{\PJ\mskip -3mu/\mskip -2mu\Ppsi}}}\xspace}
\def\psitwos  {{\ensuremath{\Ppsi{(2S)}}}\xspace}

\def\Y#1S{\ensuremath{\PUpsilon{(#1S)}}\xspace}

\def\LorLbar     {\kern \thebaroffset\optbar{\kern -\thebaroffset \PLambda}\xspace}

\def\BF         {{\ensuremath{\mathcal{B}}}\xspace}

\newcommand{\decay}[2]{\ensuremath{#1\!\to #2}\xspace} 

\def\to                 {\ensuremath{\rightarrow}\xspace}

\def\qsq       {{\ensuremath{q^2}}\xspace}

\def\CP                {{\ensuremath{C\!P}}\xspace}

\def\AT#1     {\ensuremath{A_{\mathrm{T}}^{#1}}\xspace}           

\def\C#1      {\ensuremath{\mathcal{C}_{#1}}\xspace}                       
\def\Cp#1     {\ensuremath{\mathcal{C}_{#1}^{'}}\xspace}                    
\def\Ceff#1   {\ensuremath{\mathcal{C}_{#1}^{\mathrm{(eff)}}}\xspace}        
\def\Cpeff#1  {\ensuremath{\mathcal{C}_{#1}^{'\mathrm{(eff)}}}\xspace}       
\def\Ope#1    {\ensuremath{\mathcal{O}_{#1}}\xspace}                       
\def\Opep#1   {\ensuremath{\mathcal{O}_{#1}^{'}}\xspace}                    

\newcommand{\nospaceunit}[1]{\ensuremath{\text{#1}}}       
\newcommand{\aunit}[1]{\ensuremath{\text{\,#1}}}       

\newcommand{\tev}{\aunit{Te\kern -0.1em V}\xspace}
\newcommand{\gev}{\aunit{Ge\kern -0.1em V}\xspace}
\newcommand{\mev}{\aunit{Me\kern -0.1em V}\xspace}
\newcommand{\kev}{\aunit{ke\kern -0.1em V}\xspace}
\newcommand{\ev}{\aunit{e\kern -0.1em V}\xspace}
 
\newcommand{\mevc}{\ensuremath{\aunit{Me\kern -0.1em V\!/}c}\xspace}
\newcommand{\gevc}{\ensuremath{\aunit{Ge\kern -0.1em V\!/}c}\xspace}
\newcommand{\mevcc}{\ensuremath{\aunit{Me\kern -0.1em V\!/}c^2}\xspace}
\newcommand{\gevcc}{\ensuremath{\aunit{Ge\kern -0.1em V\!/}c^2}\xspace}

\def\mum  {\ensuremath{\,\upmu\nospaceunit{m}}\xspace}

\def\fb   {\ensuremath{\aunit{fb}}\xspace}
\def\invfb   {\ensuremath{\fb^{-1}}\xspace}

\newcommand{\stat}{\aunit{(stat)}\xspace}
\newcommand{\syst}{\aunit{(syst)}\xspace}

\newcommand{\chisq}{\ensuremath{\chi^2}\xspace}

\def\gsim{{~\raise.15em\hbox{$>$}\kern-.85em
          \lower.35em\hbox{$\sim$}~}\xspace}
\def\lsim{{~\raise.15em\hbox{$<$}\kern-.85em
          \lower.35em\hbox{$\sim$}~}\xspace}

\def\pt         {\ensuremath{p_{\mathrm{T}}}\xspace}

\def\ptot       {\ensuremath{p}\xspace}

\def\evtgen     {\mbox{\textsc{EvtGen}}\xspace}

\def\geant      {\mbox{\textsc{Geant4}}\xspace}

\def\photos     {\mbox{\textsc{Photos}}\xspace}

\def\pythia     {\mbox{\textsc{Pythia}}\xspace}

\def\tell1  {TELL1\xspace}
\def\ukl1   {UKL1\xspace}

\newcommand{\lhcborcid}[1]{\href{https://orcid.org/#1}{\hspace*{0.1em}\raisebox{-0.45ex}{\includegraphics[width=1em]{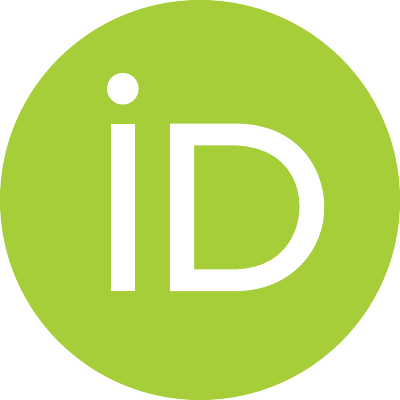}}}}

\usepackage{cite} 
\usepackage{mciteplus}

\def\Bstarz      {{\ensuremath{\B^{*0}}}\xspace}
\def\Bsstar  {{\ensuremath{\B^{*0}_\squark}}\xspace}

\def\BorBsstar  {{\ensuremath{\B^{*0}_{(\squark)}}}\xspace}

\begin{document}

\renewcommand{\thefootnote}{\fnsymbol{footnote}}
\setcounter{footnote}{1}

\begin{titlepage}
\pagenumbering{roman}

\vspace*{-1.5cm}
\centerline{\large EUROPEAN ORGANIZATION FOR NUCLEAR RESEARCH (CERN)}
\vspace*{1.5cm}
\noindent
\begin{tabular*}{\linewidth}{lc@{\extracolsep{\fill}}r@{\extracolsep{0pt}}}
\ifthenelse{\boolean{pdflatex}}
{\vspace*{-1.5cm}\mbox{\!\!\!\includegraphics[width=.14\textwidth]{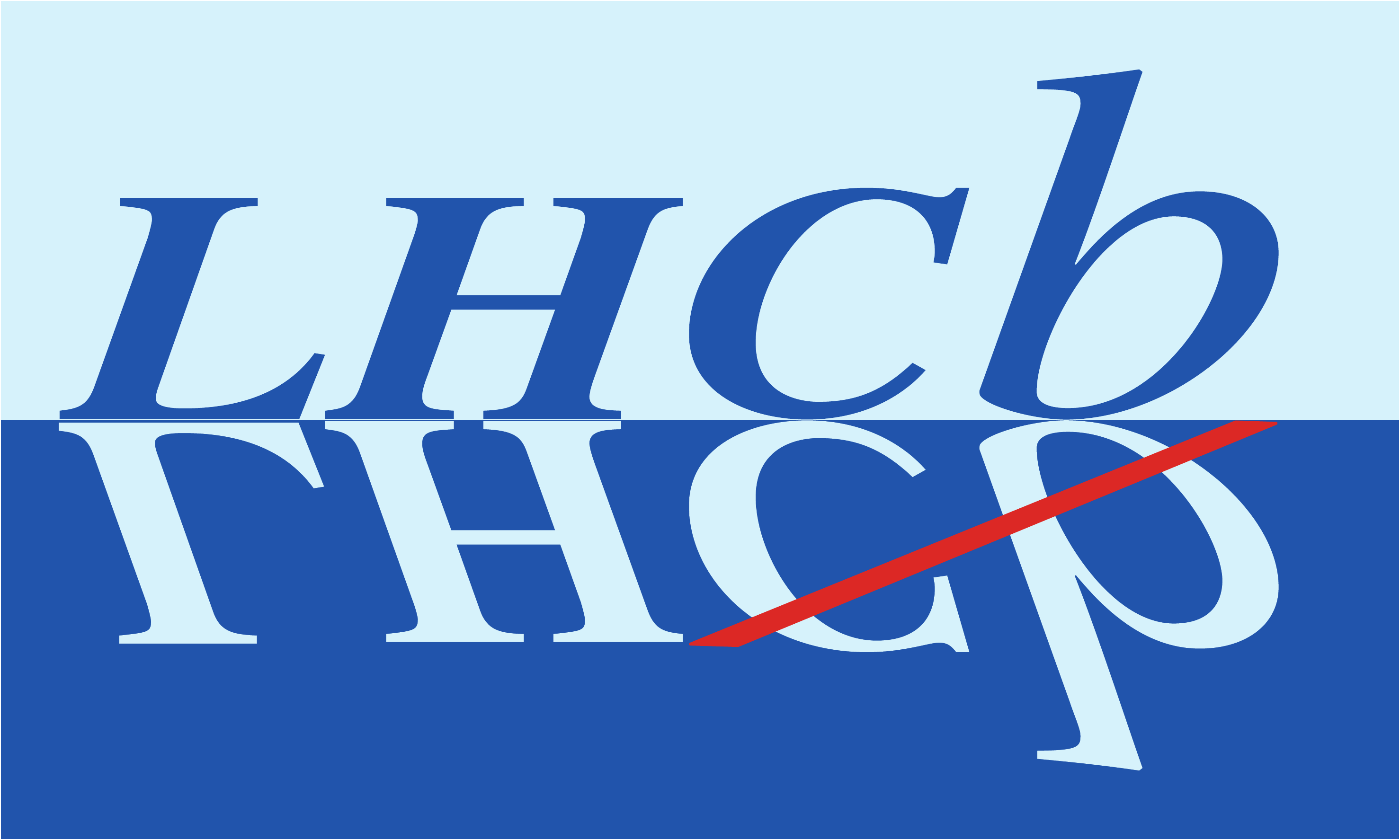}} & &}%
{\vspace*{-1.2cm}\mbox{\!\!\!\includegraphics[width=.12\textwidth]{figs/lhcb-logo.eps}} & &}%
\\
 & & CERN-EP-2023-280 \\  
 & & LHCb-PAPER-2023-037 \\  
 & & May 7, 2024 \\ 
 & & \\
\end{tabular*}

\vspace*{3.0cm}

{\normalfont\bfseries\boldmath\huge
\begin{center}
  \papertitle 
\end{center}
}

\vspace*{1.0cm}

\begin{center}
\paperauthors\footnote{Authors are listed at the end of this paper.}
\end{center}

\vspace{\fill}

\begin{abstract}
  \noindent
The first search for nonresonant \decay{\Bc}{\pip\mumu} decays is reported. 
The analysis uses proton-proton collision data collected with the LHCb detector between 2011 and 2018, corresponding to an integrated luminosity of 9\invfb.
No evidence for an excess of signal events over background is observed and an upper limit is set on the branching fraction ratio $\BF(\decay{\Bc}{\pip\mumu})/\BF(\decay{\Bc}{\jpsi\pip}) < 2.1\times 10^{-4}$ at $90\%$ confidence level.
Additionally, an updated measurement of the ratio of the \mbox{\decay{\Bc}{\psitwos\pip}} and \decay{\Bc}{\jpsi\pip} branching fractions is reported.
The ratio $\BF(\decay{\Bc}{\psitwos\pip})/\BF(\decay{\Bc}{\jpsi\pip})$ is measured to be 
$0.254\pm 0.018 \pm 0.003 \pm 0.005$, where the first uncertainty is statistical, the second systematic, and the third is due to the uncertainties on the branching fractions of the leptonic \jpsi~and \psitwos~decays. 
This measurement is the most precise to date and is consistent with previous LHCb results.
\end{abstract}

\vspace*{2.0cm}

\begin{center}
Published in \href{https://doi.org/10.1140/epjc/s10052-024-12669-x}{Eur. Phys. J. C 84, 468 (2024)}
\end{center}

\vspace{\fill}

{\footnotesize 
\centerline{\copyright~\papercopyright. \href{\paperlicenceurl}{\paperlicence}.}}
\vspace*{2mm}

\end{titlepage}


\newpage
\setcounter{page}{2}
\mbox{~}

\renewcommand{\thefootnote}{\arabic{footnote}}
\setcounter{footnote}{0}
\cleardoublepage


\pagestyle{plain} 
\setcounter{page}{1}
\pagenumbering{arabic}

\section{Introduction}
\label{sec:Introduction}

The \Bc~meson is made of the heaviest quarks that bind to form hadrons in the Standard Model~(SM): beauty and charm quarks. 
The presence of two heavy quarks distinguishes the system from other \B~mesons in theoretical calculations. 
Hence, measurements of the branching fractions of \Bc decay modes can provide unique tests of the understanding of Quantum Chromodynamics. 

The \Bc~meson can decay in various ways as each constituent quark can undergo weak decay with the other as a spectator. 
The availability of large data samples at the LHC has allowed the observation of the case where the \cquark~quark decays~\cite{LHCb-PAPER-2013-044,LHCb-PAPER-2022-034}, complementing a range of observations of transitions to final states mediated by \bquark-quark decays.
The \Bc meson can also decay through the annihilation of the \bquarkbar and \cquark~quarks into a virtual \W~boson, producing either leptonic, semileptonic or hadronic final states.
The leptonic final states are challenging to study in a hadron collider environment, because of the missing energy carried by the neutrino, but are of great interest for investigation at a future high-luminosity \epem~collider operating at the \Z~pole~\cite{Akeroyd:2008ac,Zheng:2020ult,Amhis:2021cfy,Fedele:2023gyi}.

Several \Bc~decays receiving contributions from annihilation diagrams have been searched for at LHCb, including decays into hadronic final states~\cite{LHCb-PAPER-2016-022,LHCb-PAPER-2016-023,LHCb-PAPER-2016-058} and semileptonic $\decay{\Bc}{\Dsp\mumu}$~decays~\cite{LHCb-PAPER-2022-048} (the inclusion of charge conjugate processes is implied throughout the paper). However, there has been no previous search for \Bc~decays to semileptonic final states mediated only by annihilation diagrams, such as nonresonant \decay{\Bc}{\pip\mumu} decays.
A further interest in these processes is that they may receive contributions from resonant \decay{\Bc}{\BorBsstar\pip}~decays where the vector \BorBsstar~state decays via a flavour changing neutral current transition into the \mumu final state~\cite{Abudinen:2022dnw}. 
Such \BorBsstar~decays are highly suppressed in the SM~\cite{Khodjamirian:2015dda,Grinstein:2015aua} but could be enhanced in the presence of physics beyond the SM.

In the SM, nonresonant \decay{\Bc}{\pip\mumu} decays occur through the annihilation of the \bquarkbar and \cquark~quarks into a virtual \W~boson (that decays into the \uquark and \dquarkbar~quarks that produce the charged pion) with radiation of a virtual photon or \Z~boson~(that decays into a muon pair).
Currently, there are no theoretical predictions for this decay mode. 
Weak annihilation contributions to \Bu~decays into the same final state have been investigated as these can result in interesting phenomenology including \CP-violation effects~\cite{Evans:1999kv,Beneke:2004dp,Hou:2014dza,Ali:2020tjy,Lu:2022kos}.
It is, however, unclear to what extent the theoretical approaches used in these studies are applicable to \Bc~decays.

This paper describes the first search for nonresonant \decay{\Bc}{\pip\mumu} decays exploiting proton-proton~($pp$) collision data collected with the LHCb detector, corresponding to an integrated luminosity of $9\invfb$. 
The analysis strategy is based on the reconstruction of \decay{\Bc}{\pip \mumu} candidates, sorted into intervals of the dimuon invariant mass, $m(\mumu)$, that are expected to be dominated by either nonresonant or resonant contributions.  
The intervals used are shown in Table~\ref{tab:q2ranges} (natural units with $\hbar = c = 1$ are used here and throughout the paper).
These include regions around the \jpsi\ and \psitwos\ resonances, where the \mbox{\decay{\Bc}{\jpsi\pip}} decay~mode is used as a control and normalisation channel and both the \jpsi and \psitwos resonances are reconstructed in \mumu final states.
The presence of \mbox{\decay{\Bc}{\psitwos\pip}} decays into the same final state allows an updated measurement of the ratio 
\begin{equation*}
    R_{\psitwos/\jpsi} \equiv \frac{\BF(\decay{\Bc}{\psitwos\pip})}{\BF(\decay{\Bc}{\jpsi\pip})} \,\text{,}
\end{equation*}
which has previously been measured at LHCb with smaller data samples~\cite{LHCb-PAPER-2012-054,LHCb-PAPER-2015-024}.
The $m(\mumu)$ intervals to select resonant decays correspond to about $\pm4$ times the dimuon invariant-mass resolution centred on the known resonance mass~\cite{PDG2022}.
The intervals for the nonresonant \decay{\Bc}{\pip\mumu}~decay mode are defined in terms of the dimuon mass squared, $q^2$, and include two intervals (low and central) below the \jpsi resonance, one (intermediate) between the \jpsi\ and the \psitwos\ resonances, and one (high) above the \psitwos\ resonance.
The latter lies above the production threshold for two charmed hadrons and will receive contributions from broad charmonium resonances.

\begin{table}[!tb]
    \centering
    \caption{Ranges of dimuon invariant mass or mass squared ($q^2$) used in the analysis.}    
    \begin{tabular}{l  c c}
    \toprule
     Decay mode & & Interval \\
     \midrule
       \decay{\Bc}{\pip\mumu} 
        & (low) & $0.1 < q^2 < 1.1\gev^2$ \\     
        & (central)      & $1.1 < q^2 < 8.0\gev^2$ \\
        & (intermediate)  & $11.0 < q^2 < 12.5\gev^2$\\ 
        & (high)     & $15.0 < q^2 < 35.0\gev^2$\\
        \midrule
       \multicolumn{2}{l}{\decay{\Bc}{\jpsi\,\pip}}  &  $\vert m(\mumu) - m_{\jpsi}\vert  < 50\mev$\\
       \midrule
       \multicolumn{2}{l}{\decay{\Bc}{\psitwos\,\pip}}  &  $\vert m(\mumu) - m_{\psitwos}\vert  < 50\mev$ \\ 
       \bottomrule
    \end{tabular}
    \label{tab:q2ranges}
\end{table}

For each of the intervals, a fit is performed using the \Bc candidate invariant mass, $m(\pip\mumu)$, as discriminating observable. 
The yield, relative to that for the \mbox{\decay{\Bc}{\jpsi\pip}} normalisation mode, is converted to a result on the relative branching fraction of the given $q^2$ bin, 
\begin{equation}
    R_{\pip\mumu/\jpsi\pip} \equiv \frac{{\cal B}(\decay{\Bc}{\pip\mumu})}{{\cal B}(\decay{\Bc}{\jpsi\pip})} = \frac{N_{\pip\mumu}}{N_{\jpsi\pip}} \cdot \frac{\varepsilon_{\jpsi\pip}}{\varepsilon_{\pip\mumu}} \cdot {\cal B}(\decay{\jpsi}{\mumu})\,\text{,}
\label{eq:Rpimumu}
\end{equation}
where $N$ indicates a yield, $\varepsilon$ indicates the efficiency determined from simulation with data-driven corrections, and ${\cal B}(\decay{\jpsi}{\mumu})$ is the known branching fraction of the \decay{\jpsi}{\mumu} decay~\cite{PDG2022}.
Here, ${\cal B}(\decay{\Bc}{\pip\mumu})$ indicates the $\decay{\Bc}{\pip\mumu}$ differential branching fraction integrated over the $q^2$ range relevant for the given bin.

\section{Detector and simulation} 
\label{sec:Detector} 

The \lhcb detector~\cite{LHCb-DP-2008-001,LHCb-DP-2014-002} is a single-arm forward
spectrometer covering the \mbox{pseudorapidity} range $2<\eta <5$,
designed for the study of particles containing \bquark or \cquark~quarks. 
The detector includes a high-precision tracking system
consisting of a silicon-strip vertex detector surrounding the $pp$
interaction region~\cite{LHCb-DP-2014-001}, a large-area silicon-strip detector located
upstream of a dipole magnet with a bending power of about
$4{\mathrm{\,T\,m}}$, and three stations of silicon-strip detectors and straw
drift tubes~\cite{LHCb-DP-2013-003,LHCb-DP-2017-001}
placed downstream of the magnet.
The tracking system provides a measurement of the momentum, \ptot, of charged particles with
a relative uncertainty that varies from 0.5\% at low momentum to 1.0\% at 200\gev.
The minimum distance of a track to a primary $pp$ collision vertex (PV), the impact parameter (IP), 
is measured with a resolution of $(15+29/\pt)\mum$,
where \pt is the component of the momentum transverse to the beam, in\,\gev.
Different types of charged hadrons are distinguished using information
from two ring-imaging Cherenkov detectors~\cite{LHCb-DP-2012-003}. 
Photons, electrons and hadrons are identified by a calorimeter system consisting of
scintillating-pad and preshower detectors, an electromagnetic
and a hadronic calorimeter. Muons are identified by a
system composed of alternating layers of iron and multiwire
proportional chambers~\cite{LHCb-DP-2012-002}.
The online event selection is performed by a trigger~\cite{LHCb-DP-2012-004,LHCb-DP-2019-001}, 
which consists of a hardware stage, based on information from the calorimeter and muon
systems, followed by a two-level software stage, which reconstructs the full event.

Simulation is used to optimise the event selection procedure, to model the shape of the \Bc candidate invariant-mass distributions and to estimate efficiencies accounting for the effects of the detector acceptance, reconstruction and selection criteria. 
In the simulation, $pp$ collisions are generated using \pythia~\cite{Sjostrand:2007gs} with a specific \lhcb configuration~\cite{LHCb-PROC-2010-056}. 
The production of \Bc~mesons is simulated using the dedicated generator {\textsc BcVegPy}~\cite{Chang:2005hq}.
Decays of unstable particles are described by \evtgen~\cite{Lange:2001uf}, in which final-state radiation is generated using \photos~\cite{davidson2015photos}.
The interaction of the generated particles with the detector, and its response, are implemented using the \geant toolkit~\cite{Allison:2006ve, *Agostinelli:2002hh} as described in Ref.~\cite{LHCb-PROC-2011-006}. 
The underlying $pp$ interaction is reused multiple times, with an independently generated signal decay for each~\cite{LHCb-DP-2018-004}.

The \Bc candidates reconstructed in simulation are weighted to correct for discrepancies between data and simulation associated with the particle-identification~\cite{LHCb-DP-2018-001}, track-reconstruction~\cite{LHCb-DP-2013-002} and hardware trigger~\cite{LHCb-PUB-2014-039} efficiencies.
The simulation is also corrected such that the \Bc~lifetime corresponds to the current experimental value~\cite{PDG2022,LHCb-PAPER-2013-063,LHCb-PAPER-2014-060}. 
Additional corrections are applied to account for discrepancies in \Bc production kinematics, event track multiplicity and other observables used in the selection of \Bc~candidates.
These corrections are obtained using a multivariate algorithm~\cite{Rogozhnikov:2016bdp}, which is trained using \mbox{\decay{\Bc}{\jpsi\pip}} decays in background-subtracted data and simulation. 
After the corrections are applied, the simulated distributions of all variables used in the analysis are in good agreement with the data.

\section{Candidate selection and background sources} 
\label{sec:Selection} 

Candidate $\Bc\to\pip\mumu$ decays are triggered in an identical manner as described in Ref.~\cite{LHCb-PAPER-2023-004} for \Bu\ decays to the same final states. 
The hardware stage of the trigger selects events containing at least one muon with high $\pt$. The software stage of the trigger selects events containing at least one high-\pt muon that is inconsistent with originating from any~PV. The events must contain at least one secondary vertex (formed by two or more of the final-state particles) that is significantly displaced from every PV. 
A multivariate algorithm~\cite{BBDT,LHCb-PROC-2015-018} is used to identify secondary vertices consistent with the decay of a \bquark hadron.

The initial stages of the offline selection are also similar to those for $\Bp \to \pip\mumu$ decays~\cite{LHCb-PAPER-2023-004}, except that less stringent impact parameter and flight distance requirements are imposed to account for the shorter \Bc~lifetime compared to that of the \Bp~meson. 
The \Bc~candidates are formed from pairs of well-reconstructed oppositely charged tracks that are identified as muons together with a track identified as a pion, and are required to have a good-quality vertex.
Each \Bc~candidate is associated with the PV giving the smallest value of $\chi^2_{\rm IP}$, which is defined as the difference in the vertex-fit \chisq of the PV reconstructed with and without the candidate.
Each \Bc~candidate must be consistent with originating from its associated PV, and have a momentum vector aligned with the direction between the primary and \Bc decay vertices.

Each \Bc candidate is required to have an invariant mass in the range \mbox{$6150 < m(\pip\mumu) < 6700 \mev$}. 
The expected signal resolution in $m(\pip\mumu)$ corresponds to about $20\mev$.
For the resonant \decay{\Bc}{\jpsi\pip} (\decay{\Bc}{\psitwos\pip})~decay mode, the \Bc candidate mass is calculated from a kinematic fit in which the invariant mass of the \jpsi~(\psitwos) candidate is constrained to the known \jpsi~(\psitwos)~mass~\cite{PDG2022}, thereby improving the signal resolution by a factor of about two. 

Combinatorial background arising from random combinations of tracks is suppressed using a multivariate classifier. 
A boosted decision tree~(BDT) algorithm~\cite{Breiman}, as implemented in the \texttt{XGBoost} library~\cite{Chen:2016:XST:2939672.2939785}, is trained to identify the \decay{\Bc}{\pip\mumu} signal candidates irrespective of dimuon invariant mass. 
The signal sample used for the training comprises simulated nonresonant \decay{\Bc}{\pip\mumu} decays, resonant \decay{\Bc}{\jpsi\pip} decays, and \decay{\Bc}{\Bstarz(\mumu)\pip} and \decay{\Bc}{\Bsstar(\mumu)\pip} decays. 
The latter two populate the high \qsq\ region, which is sparsely populated in the nonresonant \decay{\Bc}{\pip\mumu} simulation.
The background training sample comprises data from low- and high-mass sideband regions corresponding to $5550 < m(\pip\mumu) < 5850 \mev$ and $6700 < m(\pim\mumu) < 7000\mev$, with vetoes to remove candidates containing \decay{\jpsi}{\mumu} and \decay{\psitwos}{\mumu} decays. 
The features of the data used to classify \Bc~candidates are: the \pt of the pion track; the \pt of the muon track with highest \pt; the IPs of the muon tracks and the \Bc~candidate, the \Bc flight distance, the vertex quality of the \Bc~candidate; and the largest distance of closest approach between any two of the final-state particles. 

Requirements on the BDT output and variables characterising the charged pion particle identification are optimised simultaneously using a grid search to obtain the best signal sensitivity. 
The requirements are optimised independently for the search for the nonresonant \decay{\Bc}{\pip\mumu} decays and for the measurement of $R_{\psitwos/\jpsi}$.  
For the search for the \decay{\Bc}{\pip\mumu} decay, the optimisation is based on the figure of merit $\varepsilon / (5/2 + \sqrt{N_{\rm B}})$~\cite{Punzi:2003bu}, where $\varepsilon$ is the efficiency on the combined signal sample and $N_{\rm B}$ is the expected number of background candidates in the signal region (corresponding to an interval of about $\pm3$ times the \Bc~invariant-mass resolution centred on the known \Bc~mass~\cite{PDG2022}).
The expected background yield is estimated by fitting a background-only model to the dataset excluding the signal region. 
For the measurement of $R_{\psitwos/\jpsi}$, the optimisation is based on the figure of merit $N_{\rm S}/\sqrt{N_{\rm S}+N_{\rm B}}$, where $N_{\rm S}$ and $N_{\rm B}$ correspond to the expected number of signal and background events in the signal region for the \decay{\Bc}{\psitwos\pip} decay. 
The expectation for $N_{\rm S}$ is calculated from the signal efficiency and the \decay{\Bc}{\psitwos\pip} decay yield from a fit to the data with only a loose requirement on the BDT output applied. The fit model is described below. 

The optimal requirements for the nonresonant \decay{\Bc}{\pip\mumu} search and for the $R_{\psitwos/\jpsi}$ measurement are referred to subsequently as the $\pip\mumu$ and the $\psitwos\pip$ working points~(WPs), respectively.
At the $\pip\mumu$ WP, the classifier has a combinatorial background rejection power of $99.9\%$, whilst retaining $27.1\%$ of signal decays.
At the $\psitwos\pip$ WP, the classifier has a combinatorial background rejection power of $98.8\%$, whilst retaining $63.3\%$ of signal decays. 
In both cases, the particle identification requirements have a pion efficiency around $90\%$, with a kaon misidentification rate around $10\%$.  
The particle identification requirements applied on the muon candidates have an efficiency around $99\%$, with a pion misidentification rate below $1\%$. 
To measure branching fraction ratios, the same selection is used for the signal and for the normalisation modes, such that potential systematic biases on the measurement are reduced. 
For both WPs, each selected event contains only one retained candidate. 

Backgrounds from partially reconstructed decays such as \mbox{\decay{\Bc}{\rhop\mumu}} (with \rhop decaying into $\pip\piz$) for the nonresonant signal mode, \decay{\Bc}{\jpsi\rhop} for the normalisation mode, or \decay{\Bc}{\psitwos\rhop} for the signal \decay{\Bc}{\psitwos\pip} mode, have a reconstructed \Bc candidate invariant mass that lies more than $100\mev$ below the known \Bc~mass~\cite{PDG2022}. 
These sources of background predominantly populate a region outside, but have a tail that extends into, the fit range used in the analysis. 
This contribution is neglected in the fits to the nonresonant \decay{\Bc}{\pip\mumu} candidates, but is accounted for in the resonant mode fits.
Backgrounds with a missing neutrino or two or more missing massive particles can also be a source of partially reconstructed background, but their contributions are negligible in the fit range. 

Due to the imperfect pion-kaon separation, misidentified backgrounds can in principle arise from \decay{\Bc}{\Kp\mumu} decays for the nonresonant signal mode, \decay{\Bc}{\jpsi\Kp} for the \decay{\Bc}{\jpsi\pip} mode, or \decay{\Bc}{\psitwos\Kp} for the \decay{\Bc}{\psitwos\pip} mode.
However, the branching fractions for these sources of background are Cabibbo-suppressed with respect to the signal decays. 
For instance, the ratio of the \decay{\Bc}{\jpsi\Kp} and \decay{\Bc}{\jpsi\pip} branching fractions has been measured to be $0.079\pm 0.007 \pm 0.003$~\cite{LHCb-PAPER-2016-020}.
These background contributions are further suppressed by the particle identification requirements. 
Due to its estimated size, the Cabibbo-suppressed background is neglected for the search for \decay{\Bc}{\pip\mumu} decays.
The corresponding background sources are, however, included in the resonant signal mode fits.

Contributions from hadronic backgrounds such as \mbox{\decay{\Bc}{\pip\pim\pip}} decays, where two pions are mistakenly identified as muons are expected to be negligible based on the misidentification rates estimated from simulation and the fact that these decays --- which have not yet been observed --- are expected to occur via suppressed annihilation processes. 
For the search for the nonresonant \decay{\Bc}{\pip\mumu} decay, it is in principle possible to receive contributions from the resonant \decay{\Bc}{\jpsi\pip} or \decay{\Bc}{\psitwos\pip} decays, where the \pip is mistakenly identified as a \mup and vice versa. These sources of background are studied using simulation and data and found to be negligible after applying the selection requirements.    

\section{Invariant-mass fits} 
\label{sec:Fit} 

The yields for nonresonant \decay{\Bc}{\pip\mumu} decays, resonant \decay{\Bc}{\jpsi\pip} decays, and resonant \decay{\Bc}{\psitwos\pip} decays are determined from one-dimensional extended unbinned maximum-likelihood fits to the $m(\pip\mumu)$ distribution of the selected candidates in the respective dimuon invariant-mass intervals~(see Table~\ref{tab:q2ranges}). 
For the resonant decays, the fits are performed independently to the \Bc~candidates fulfilling the $\pip\mumu$ WP requirements and for \Bc~candidates fulfilling the $\psitwos\pip$ WP requirements~(see Sec.~\ref{sec:Selection}). 
The fits at the $\pip\mumu$ WP are performed to the $\pip\mumu$ invariant-mass distribution without additional constraints. 
The fits at the $\psitwos\pip$ WP are performed to the \Bc candidate mass distribution after constraining the dimuon mass to the known \jpsi\ or \psitwos\ mass~\cite{PDG2022}, as appropriate. 

For the search for the nonresonant \decay{\Bc}{\pip\mumu} decay, fits are performed independently for each of the four $q^2$~intervals and for the combination of the four intervals. 
The fit model includes two components: signal decays and combinatorial background.
The signal model is validated using simulation and corresponds to the sum of two Gaussian functions, one with power-law tails on both sides of the distribution~\cite{Skwarnicki:1986xj}.
The tail parameters are fixed to the values obtained from simulation. 
The signal model includes a global shift of peak position and a global scaling factor for the width of the distribution, relative to the values found in simulation.
The peak offset and width scale factor are obtained from a fit to the resonant \decay{\Bc}{\jpsi\pip} decays in data.
The combinatorial background is described by an exponential function, whose exponent is allowed to vary in the fit to data.  

\begin{figure}[!tb]
    \centering

    \begin{tabular}{cc}
    \includegraphics[width=0.49\linewidth]{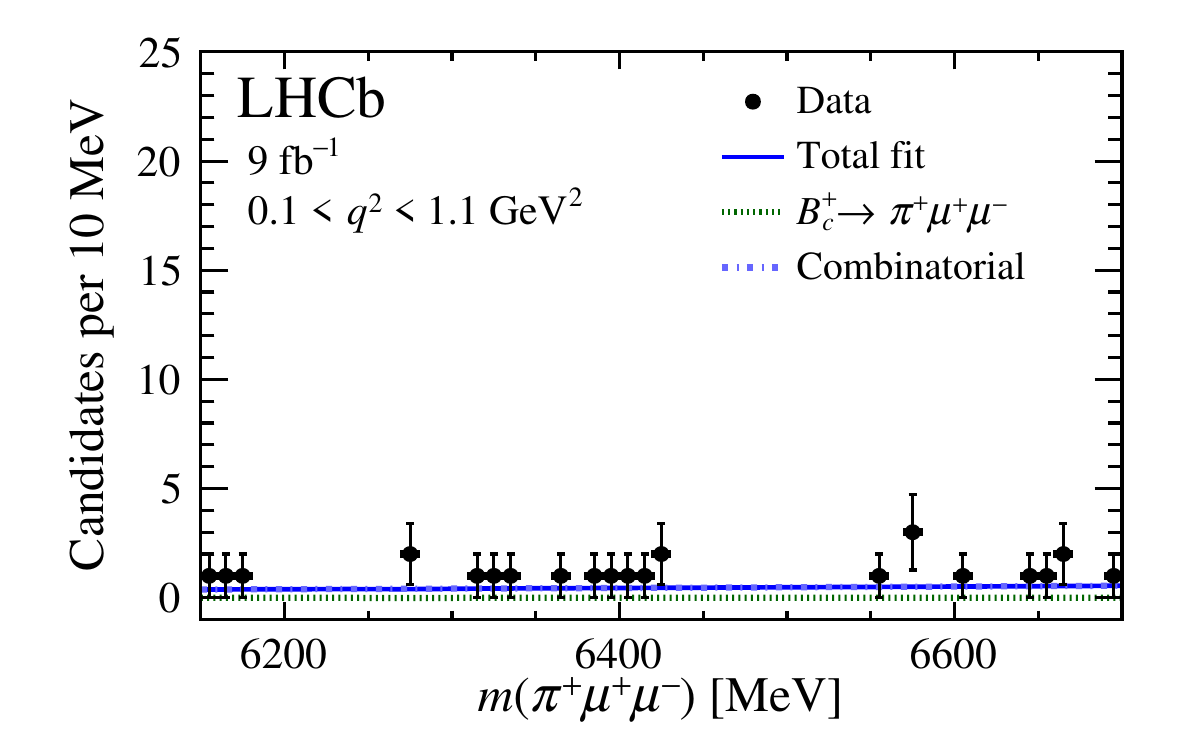} &
    \includegraphics[width=0.49\linewidth]{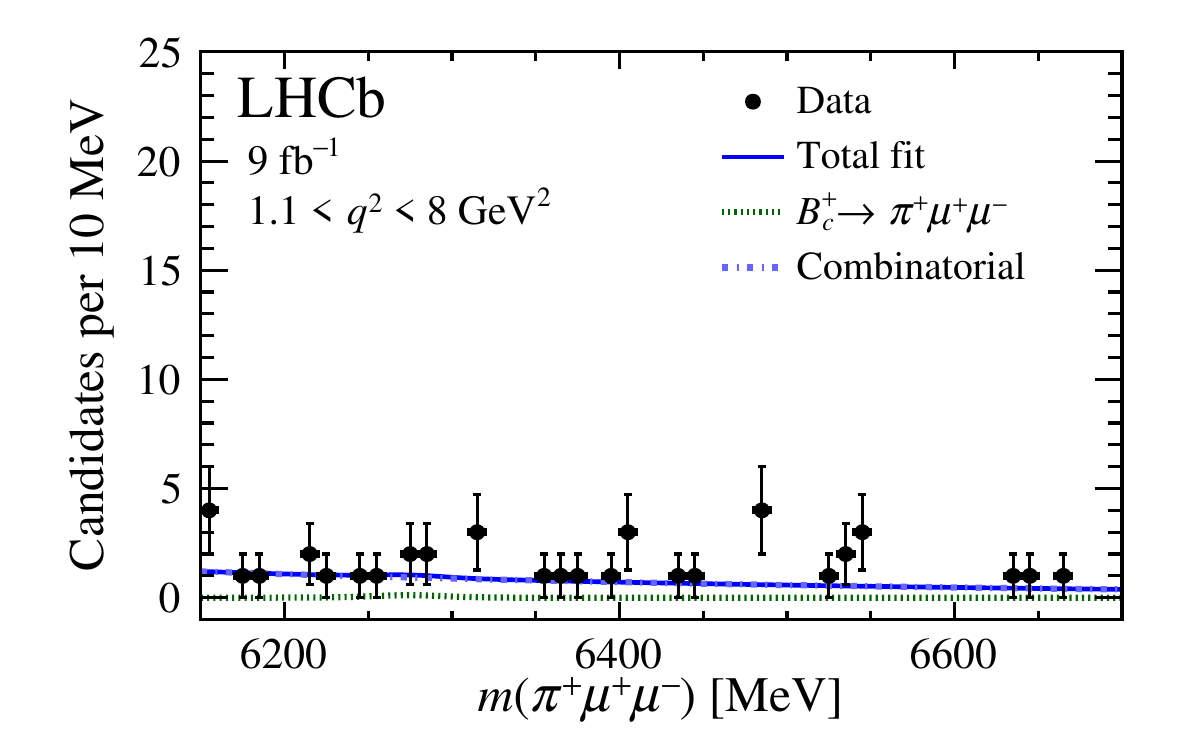} \\
    Low & Central \\

    \includegraphics[width=0.49\linewidth]{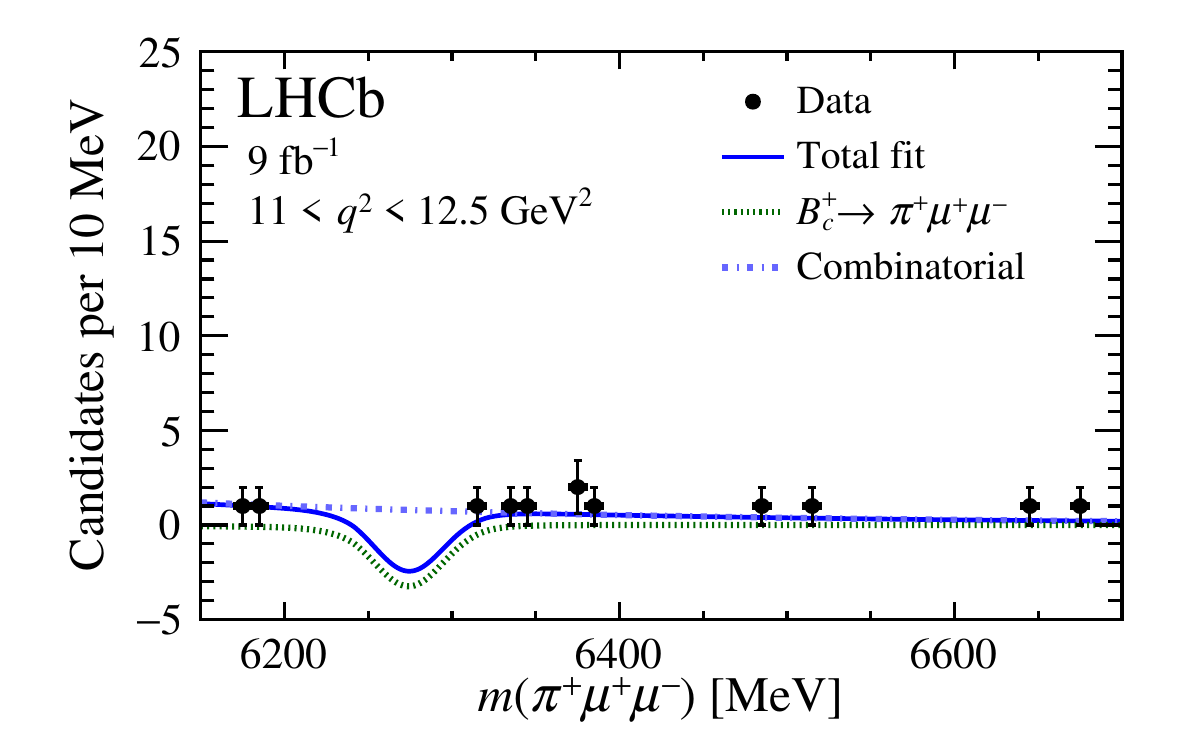} &
    \includegraphics[width=0.49\linewidth]{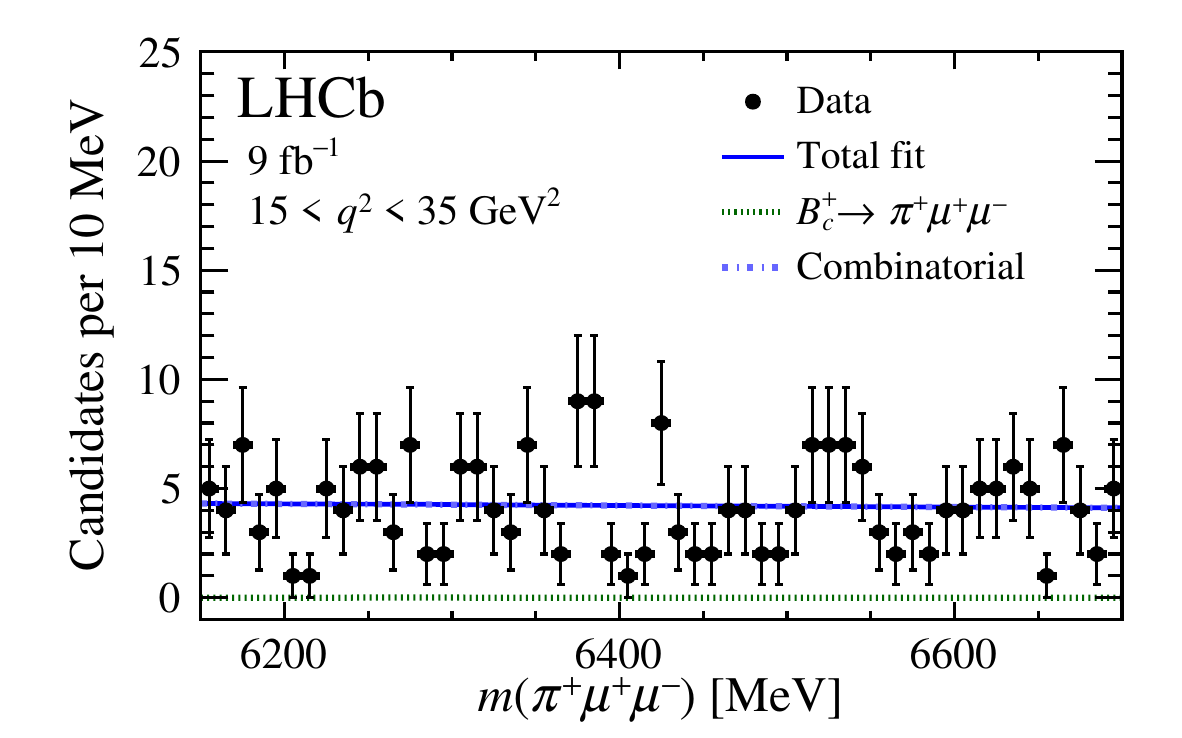} \\
    Intermediate & High \\

    \multicolumn{2}{c}{\includegraphics[width=0.49\linewidth]{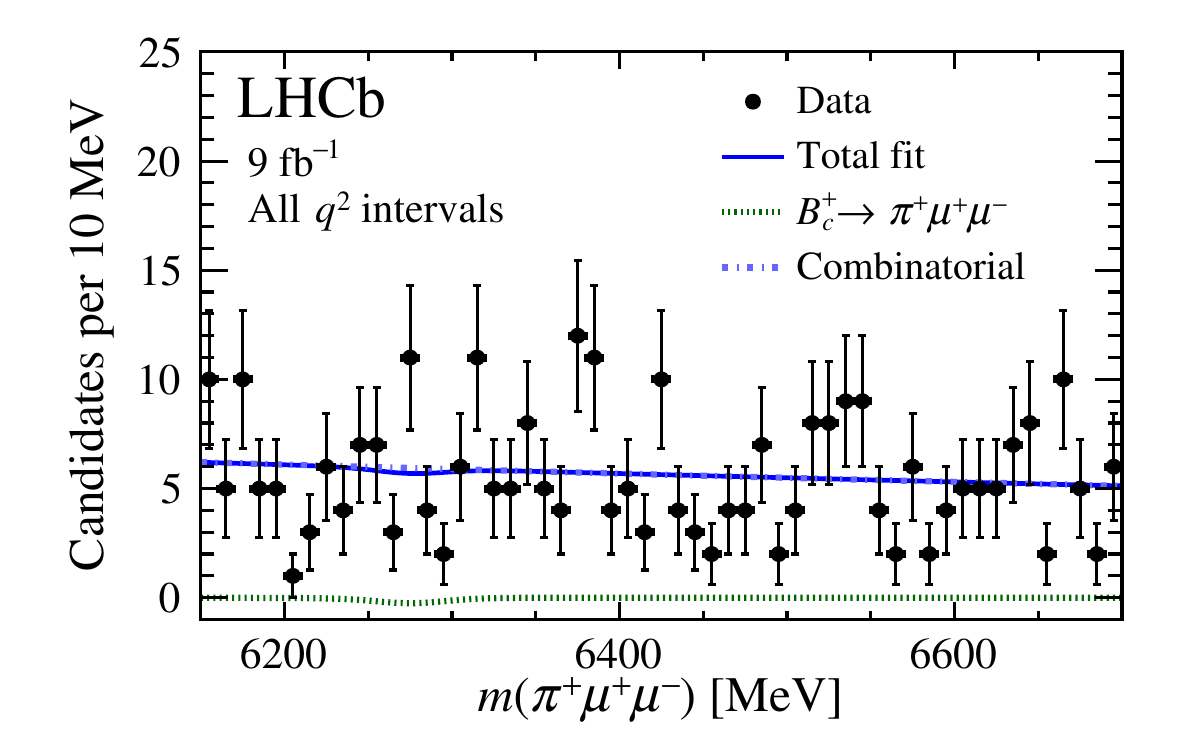}} \\
    \multicolumn{2}{c}{All}
    \end{tabular}
    \caption{Reconstructed $\pip\mumu$~invariant-mass distributions for the selected \decay{\Bc}{\pip\mumu} candidates for each $q^2$~interval and for all intervals combined, with results of the fit described in the text overlaid. }
    \label{fig:unblinded_fits}
\end{figure}

Figure~\ref{fig:unblinded_fits} shows the \Bc candidate invariant-mass distributions of selected \mbox{\decay{\Bc}{\pip\mumu}} candidates, with results of the fits superimposed.
For the intermediate $q^2$~interval, the fit favours a negative signal yield due to a lack of candidates in the region close to the \Bc~mass peak position and a feature of extended maximum-likelihood fits that prefers negative yields in such situations~\cite{porter-book}. 
Table~\ref{tab:fitresults_fixed} summarises the yields obtained from the fits.

\begin{table}[tb!]
    \renewcommand{\arraystretch}{1.2}
    \centering
    \caption{Yields for the signal nonresonant \decay{\Bc}{\pip\mumu}~decay~($N_{\pip\mumu}$) and combinatorial background~($N_{\rm comb}$) obtained from the fits to data described in the text, with statistical uncertainties only. }    
 \begin{tabular}{r@{$\,< q^2 <\,$}l c c}
 \toprule
   \multicolumn{2}{c}{$q^2$ interval}   &  
   $N_{\pip\mumu}$ &
   $N_{\rm comb}$ \\
    \midrule
    0.1 & $\phantom{1}1.1\gev^2$    &   $\phantom{{-1}}0 \pm 2$          & $25\,^{+6}_{-5}$    \\ 
    1.1 & $\phantom{1}8.0\gev^2$    &   $\phantom{{-1}}1\,^{+4}_{-3}$    & $39 \pm 7$    \\ 
   11.0 & $12.5\gev^2$   &   $-18\,^{+7}_{-10}$         & $\;\,30\,^{+13}_{-9}$   \\ 
   15.0 & $35.0\gev^2$   &   $\phantom{{-1}}0\,^{+8}_{-7}$    & $232 \pm 17$ \\ 
\multicolumn{2}{c}{All} &   $\phantom{1}{-2}\,^{+9}_{-8}$      & $311\,^{+20}_{-19}$ \\
\bottomrule
 \end{tabular}
\label{tab:fitresults_fixed}
\end{table}

For the resonant \decay{\Bc}{\jpsi\pip} decay, the fit model includes four components: signal \decay{\Bc}{\jpsi \pip} decays, misidentified \decay{\Bc}{\jpsi \Kp} decays, partially reconstructed background from \decay{\Bc}{\jpsi \rhop} decays and combinatorial background. 
The signal, misidentified and partially reconstructed backgrounds are each parameterised by the sum of two Gaussian functions, one with power-law tails.  The tail parameters of each distribution are fixed from simulation. 
The peak position and width of the distributions are allowed to vary in the fit to the data by a global offset and scale factor that is shared between the components. 
The combinatorial background model is an exponential function, whose exponent is allowed to vary in the fit to data.
In total, the \decay{\Bc}{\jpsi\pip} fit includes seven free parameters: the yields of the four components, the global peak position shift and width scaling factor, and the exponent of the combinatorial background. 

\begin{figure}[!bt]
    \centering
    \includegraphics[width=0.49\textwidth]{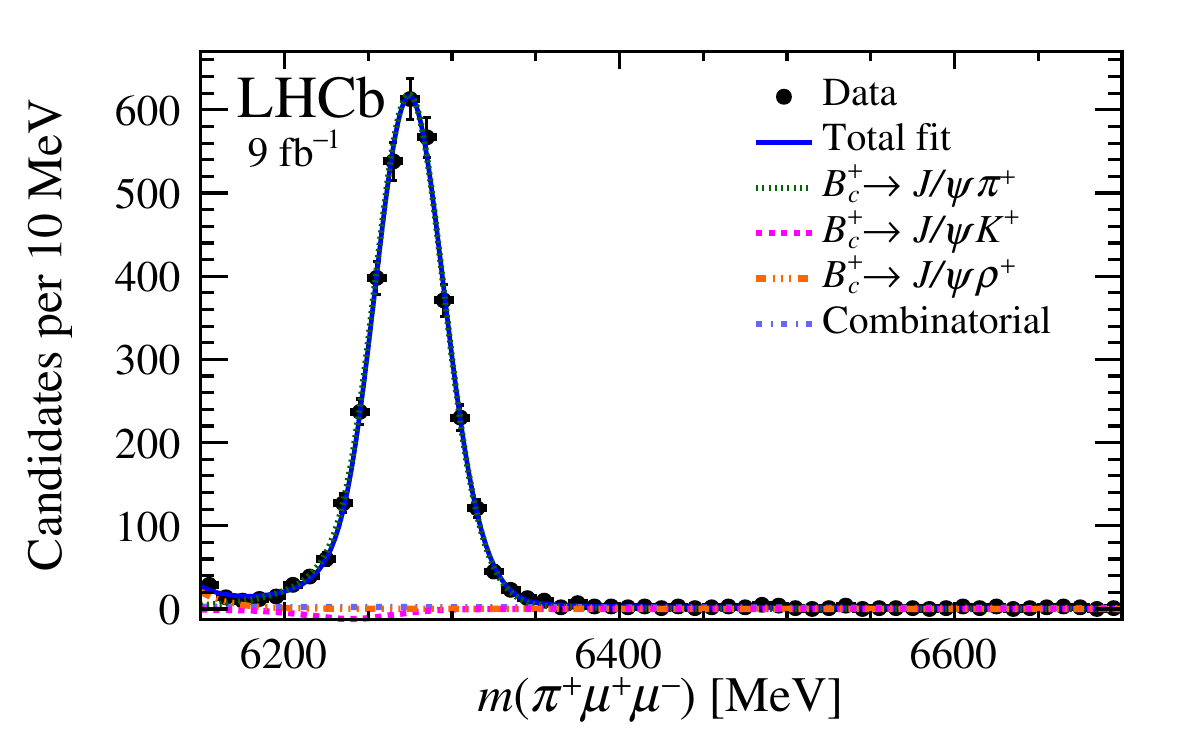}
    \includegraphics[width=0.49\textwidth]{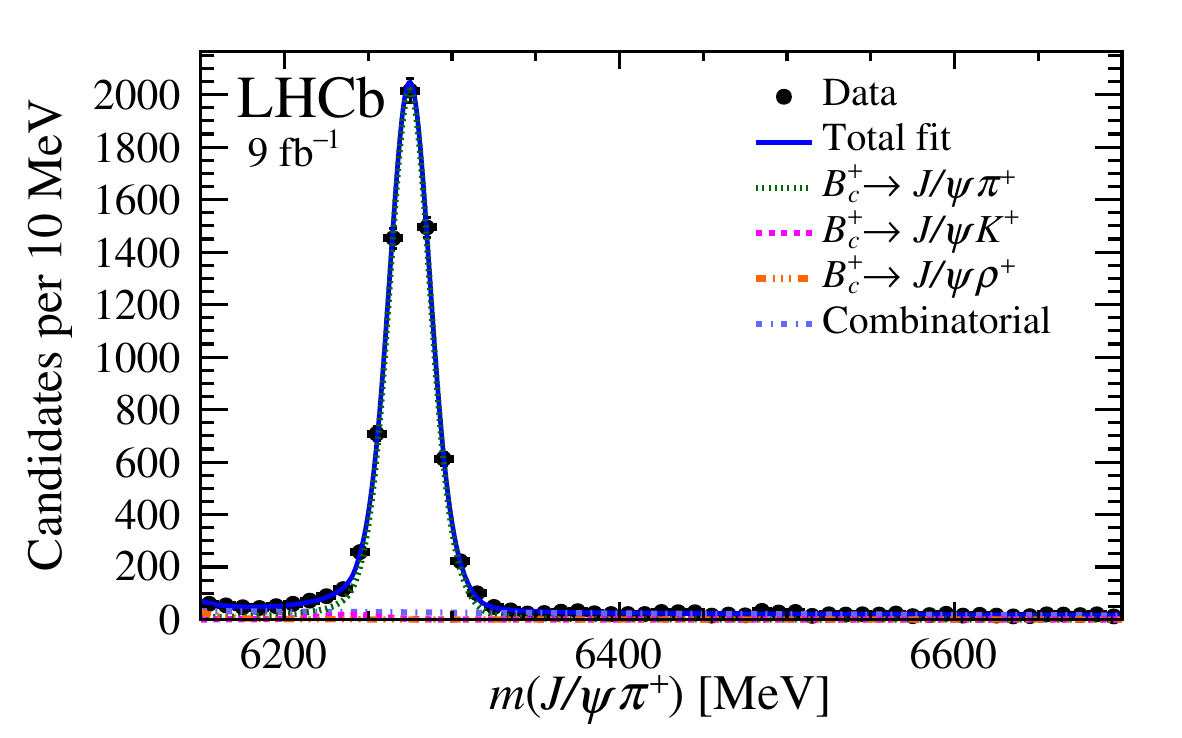}
    \caption{Reconstructed $\pip\mumu$ invariant-mass distributions for the selected \mbox{\decay{\Bc}{\jpsi\pip}} candidates at the (left) $\pip\mumu$ and (right) $\psitwos\pip$ working points, with the results of the fits overlaid. 
    For the right figure, the $\pip\mumu$ invariant mass is calculated after constraining the dimuon mass to the known \jpsi~mass, thereby improving the signal resolution.
    }
    \label{fig:jpsipi_fit}
\end{figure}

Figure~\ref{fig:jpsipi_fit} shows the distributions of selected \mbox{\decay{\Bc}{\jpsi\pip}} candidates for both WPs, with fit projections overlaid. 
Table~\ref{tab:jpsipi_data_fit_params} summarises the yields obtained from the fits.
The ratio of \decay{\Bc}{\jpsi \pip}~yields at the two WPs is found to be consistent with the expectation based on the efficiencies in simulation.
The yields of the misidentified \decay{\Bc}{\jpsi \Kp}~component at the two WPs are found to be consistent with the expectations based on the misidentification rates in simulation and the measured ratio of branching fractions $\BF(\decay{\Bc}{\jpsi\Kp})/\BF(\decay{\Bc}{\jpsi\pip})$~\cite{LHCb-PAPER-2016-020}.

\begin{table}[tb!]
    \centering
    \caption{Yields obtained from the \decay{\Bc}{\jpsi\pip} fits to data at the two working points as described in the text, with statistical uncertainties only. }
    \begin{tabular}{lr@{$\,\pm\,$}lr@{$\,\pm\,$}l}    
        \toprule
        Component & \multicolumn{2}{c}{$\pip\mumu$ WP} & \multicolumn{2}{c}{$\psitwos\pip$ WP} \\
        \midrule
 \decay{\Bc}{\jpsi\pip}         & \quad$3508$ & $82$        & \quad$6887$ & $93$\\
 \decay{\Bc}{\jpsi\Kp}          & $-81$  & $58$        & $90$   & $43$\\
 \decay{\Bc}{\jpsi\rhop}        & $41$   & $11$        & $56$   & $22$\\
 Comb.\ bkg.                    & $101$  & $25$        & $1254$ & $60$ \\
 \bottomrule 
    \end{tabular}
    \label{tab:jpsipi_data_fit_params}
\end{table}

\begin{figure}[!bt]
    \centering
    \includegraphics[width=0.49\textwidth]{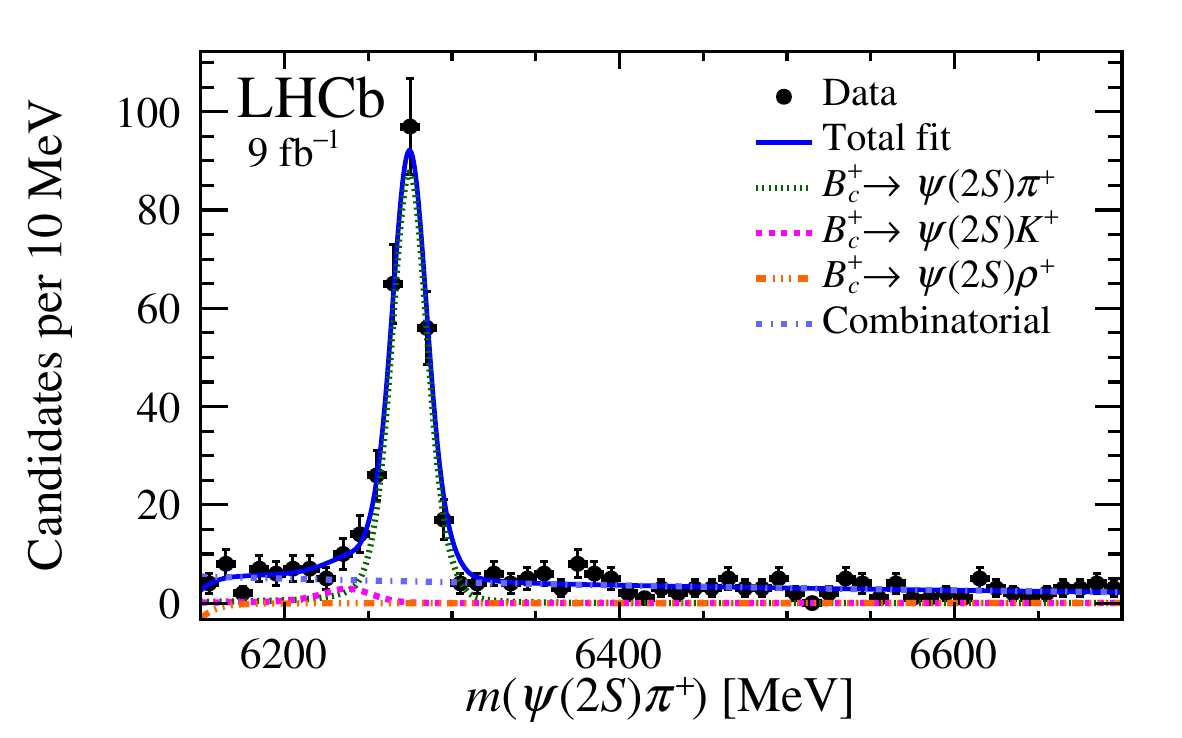}
    \caption{Reconstructed $\pip\mumu$ invariant-mass distribution, calculated after constraining the dimuon mass to the known \psitwos~mass, for the selected \mbox{\decay{\Bc}{\psitwos\pip}} candidates, with results of the fit described in the text overlaid.}
    \label{fig:psitwospi_fit}
\end{figure}

Similarly, for the resonant \decay{\Bc}{\psitwos\pip} decay, the fit includes four components: signal \decay{\Bc}{\psitwos \pip} decays, misidentified background from \decay{\Bc}{\psitwos \Kp} decays, partially reconstructed \decay{\Bc}{\psitwos \rhop} background and combinatorial background. 
The analytical functions of the fit models used for the \decay{\Bc}{\psitwos\pip} fit are the same as those used for the \decay{\Bc}{\jpsi\pip} fit, but with different parameters. 
The signal model, and the models for the misidentified and the partially reconstructed background have tail parameters which are fixed to values obtained from simulation, and they also include a global shift of peak position and a global scaling factor for the widths of the distributions.
However, the global peak position shift and width scaling factor are constrained in the fit to data to be consistent with the values obtained from the \decay{\Bc}{\jpsi\pip} fit.
Thus, the \decay{\Bc}{\psitwos\pip} fit includes five free parameters: the yields for the four components and the exponent of the combinatorial background. 
The \decay{\Bc}{\psitwos\pip} fit is performed to the \Bc mass distribution after constraining the dimuon mass to the known \psitwos~mass. 
Figure~\ref{fig:psitwospi_fit} shows the distribution of selected \decay{\Bc}{\psitwos\pip} candidates, with fit projections overlaid. 
Table~\ref{tab:psitwos_data_fit_params} summarises the yields obtained from the fit.

\begin{table}[tb!]
    \centering
    \caption{Yields obtained from the \decay{\Bc}{\psitwos\pip} fit to data described in the text, with statistical uncertainties only. }
    \begin{tabular}{lr@{$\,\pm\,$}l}    
        \toprule
        Component & \multicolumn{2}{c}{Yield} \\ 
        \midrule
 \decay{\Bc}{\psitwos\pip}           & $256$ & $18$ \\
 \decay{\Bc}{\psitwos\Kp}            & $13$  & $10$ \\
 \decay{\Bc}{\psitwos\rhop}          & $-4$  & $5$  \\
 Comb.\ bkg.                         & $197$ & $19$ \\ 
 \bottomrule
    \end{tabular} 
    \label{tab:psitwos_data_fit_params}
\end{table}

\section{Efficiencies and systematic uncertainties} 
\label{sec:Systemtics} 

The efficiency ratios between signal and normalisation modes are obtained from simulation accounting for the geometrical acceptance of the detector as well as effects related to the triggering, reconstruction and selection of the \Bc~candidates.
Table~\ref{tab:effcies} lists the efficiency ratios between signal and normalisation modes.
For the nonresonant signal mode, the efficiency denominator includes only the generated events in the respective $q^2$ interval.
The uncertainties on the efficiency ratios take into account the simulation sample size, uncertainties on the weights applied to the simulation, 
the matching between reconstructed and generated particles in the simulation,
variations of the software trigger requirements,
and the uncertainty on the known \Bc\ lifetime. 
All variations are made consistently for the signal and normalisation modes to avoid overestimation of the uncertainty on the efficiency ratio. 
For the nonresonant signal mode, the impact of the signal decay model assumed in the simulation is additionally considered. 

\begin{table}[!tb]
    \centering
    \caption{Efficiency ratios between signal and normalisation modes. The values of $\varepsilon_{\jpsi\pip}/\varepsilon_{\pip\mumu}$ are provided for each $q^2$ interval and for all intervals combined. The uncertainty on the ratio combines statistical and systematic effects. }
    \begin{tabular}{l c r@{$\,\pm\,$}l}
    \toprule
     Signal decay mode   &  &    \multicolumn{2}{c}{ $\varepsilon_{\jpsi\pip}/\varepsilon_{\pip\mumu}$} \\
     \midrule
       \decay{\Bc}{\pip\mumu}
        & (low)   & \quad1.26 & 0.24\\     
        & (central)   & 1.27 & 0.18\\
        & (intermediate) & 0.91 & 0.04\\ 
        & (high) & 0.86 & 0.24\\
        & (all)                      & 1.01 & 0.27\\\hline
       \multicolumn{2}{l}{\decay{\Bc}{\psitwos\pip}}  & 0.91  & 0.01
       \\
       \bottomrule
    \end{tabular}
    \label{tab:effcies}
\end{table}

The systematic uncertainties associated with the weights are evaluated by varying all weights within their uncertainties and by varying the binning scheme used to estimate them. 
The systematic uncertainty associated with the multivariate weighting algorithm~(see Sec.~\ref{sec:Detector}) is evaluated by comparing the results obtained with the default and with an alternative algorithm. 
The default algorithm is trained to correct for discrepancies between data and simulation associated with the event track multiplicity and with the \pt and the vertex quality of the \Bc~candidates.
The alternative algorithm is trained using the IPs of the two muons as additional inputs. 

The systematic uncertainty associated with the matching between reconstructed and generated particles in the simulation is evaluated by comparing the efficiencies obtained including or excluding \B~candidates for which one or more decay products are not correctly matched. 
The systematic uncertainty associated with variations of the software trigger requirements that are not mirrored by the simulation is evaluated by comparing the efficiencies obtained by applying the tightest thresholds and by applying average thresholds within each data-taking period.  
The systematic uncertainty associated with the \Bc\ lifetime is evaluated by varying the \Bc\ lifetime in simulation within its uncertainties~\cite{PDG2022}.

\begin{figure}[tb]
  \centering
    \includegraphics[width=0.7\linewidth]{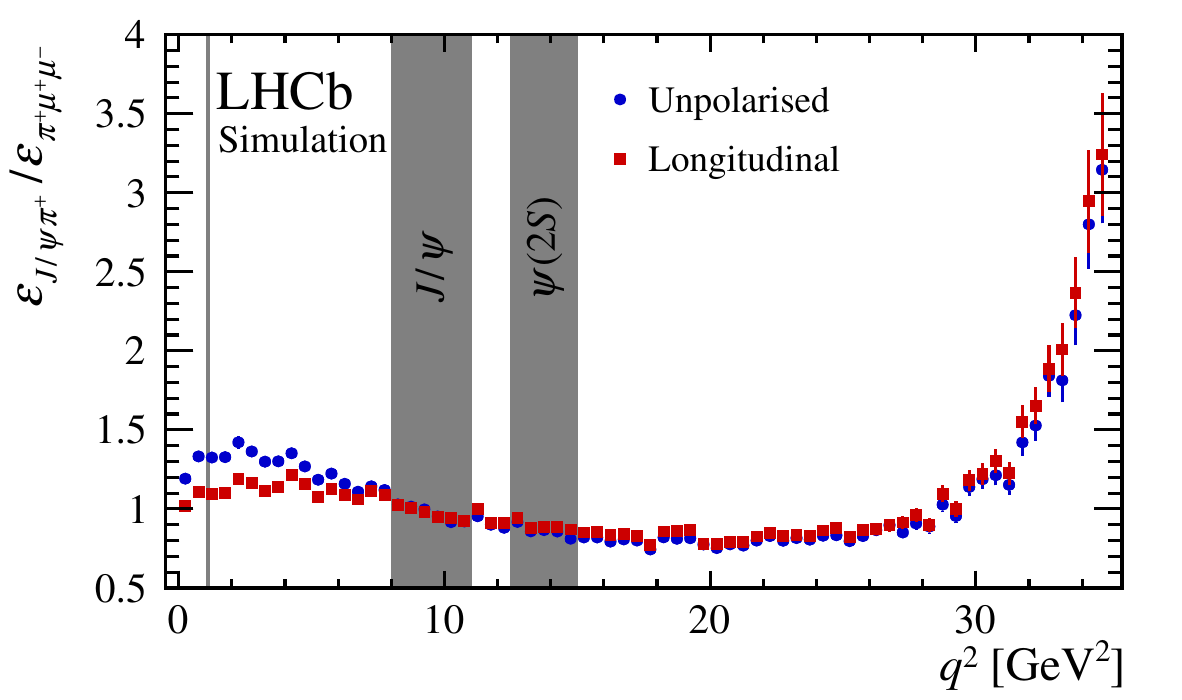}
  \caption{
  \label{fig:effcy_q2_var_SAME}
    Efficiency ratio between the \decay{\Bc}{\jpsi\pip} and nonresonant \decay{\Bc}{\pip\mumu} decay modes as a function of $q^2$, evaluated for the two extreme possibilities of the dimuon system forming a scalar state, which is unpolarised, and a vector state that is longitudinally polarised.  
    The shaded $q^2$ intervals, which contain the contributions from the \jpsi\ and \psitwos\ resonances, are not used in the analysis of nonresonant decays.
  }
\end{figure}

The nonresonant signal decays are simulated with a phase-space distribution of the final-state particles. 
However, the true $q^2$ distribution of $\Bc \to \pip\mumu$ decays is unknown, and there is no clear theoretical guidance for its expectation.
A systematic uncertainty is assigned due to the change of efficiency over $q^2$, shown in Fig.~\ref{fig:effcy_q2_var_SAME}.
The uncertainty is taken to be the root-mean-square variation of the distribution of the efficiency values in narrow $q^2$ bins within each interval.
The approach of determining the differential branching fraction within $q^2$ bins helps to reduce this uncertainty.

A similar uncertainty arises due to the different possible polarisations in the dimuon system, and is illustrated by the different coloured distributions in Fig.~\ref{fig:effcy_q2_var_SAME}. 
The two extreme possibilities correspond to the dimuon system forming a scalar state, which is unpolarised, and a vector state that is longitudinally polarised.  
The two cases are characterised by different $\cos{\theta_l}$ distributions, where the helicity angle $\theta_l$ is defined as the angle between the \mup~momentum and the opposite of the \Bp~momentum in the dimuon frame: the scalar dimuon state corresponds to a flat $\cos{\theta_l}$ distribution, while the vector dimuon state corresponds to a $\frac{3}{4}\sin^2 {\theta_l}$ distribution.
The difference in efficiency between the two extreme cases is taken as the associated systematic uncertainty.

For the resonant modes, the effect of the multivariate weighting algorithm dominates the systematic uncertainty on the efficiency ratio, while for the nonresonant signal mode, the systematic uncertainty associated with efficiency variation within the $q^2$ intervals dominates the uncertainty on the efficiency ratio. 
The remaining systematic uncertainties cancel out almost fully in the determination of the efficiency ratios. 
For all measured quantities, the systematic uncertainties are smaller than the statistical uncertainties.  

The yields obtained from the fits described in the previous section can be affected by the fit model choice and by the assumption of the polarisation of the partially reconstructed backgrounds.
To study the effect of the fit model choice, each fit is performed in two configurations: using the baseline fit model and using an alternative fit model. 
In the alternative fit model, the analytical function used for the signal and the misidentified background models is replaced with the sum of two Gaussian functions, one with a power-law tail on the left side of the distribution, and the other with a power-law tail on the right.
The analytical function for the partially reconstructed backgrounds is also replaced, but with the sum of two Gaussian functions.
In all cases the difference between the results obtained with the baseline and alternative models is found to be negligible.

For the fits to nonresonant signal candidates, the possible contribution from partially reconstructed \decay{\Bc}{\rhop\mumu} decays is found to be negligible by performing fits to sideband data of background-only models including a \decay{\Bc}{\rhop\mumu} background component.
The sideband data includes \Bc~candidates in the fit range excluding those in the signal region~(see Sec.~\ref{sec:Selection}).
For the resonant modes, the \rhop~meson is assumed to be unpolarised. However, the polarisation of the \rhop~meson can affect the momentum of the missing pion and hence the \Bc candidate mass shape of the partially reconstructed backgrounds.
To study the effect of the \rhop~polarisation, the fits are repeated assuming either full longitudinal or full transverse \rhop~polarisation. 
The difference in the results for the two configurations is found to be negligible.

\section{Results} 
\label{sec:Result} 

For the nonresonant signal mode, no signal is observed over the background-only hypothesis and an upper limit is set on $R_{\pip\mumu/\jpsi\pip}$ in each $q^2$ bin.
The ratio $R_{\pip\mumu/\jpsi\pip}$ is obtained by parametrising the signal yield $N_{\pip\mumu}$ in the fits described in Sec.~\ref{sec:Fit} using Eq.~\eqref{eq:Rpimumu}. 
The branching fraction $\BF\left(\decay{\jpsi}{\mumu}\right)$~\cite{PDG2022}, the efficiency ratio $\varepsilon_{\jpsi\pip}/\varepsilon_{\pip\mumu}$, and the normalisation yield $N_{\jpsi\pip}$ are allowed to vary within Gaussian constraints to account for the uncertainties on these inputs.  

\begin{figure}[!htb]
    \centering

    \begin{tabular}{cc}
    \includegraphics[width=0.49\linewidth]{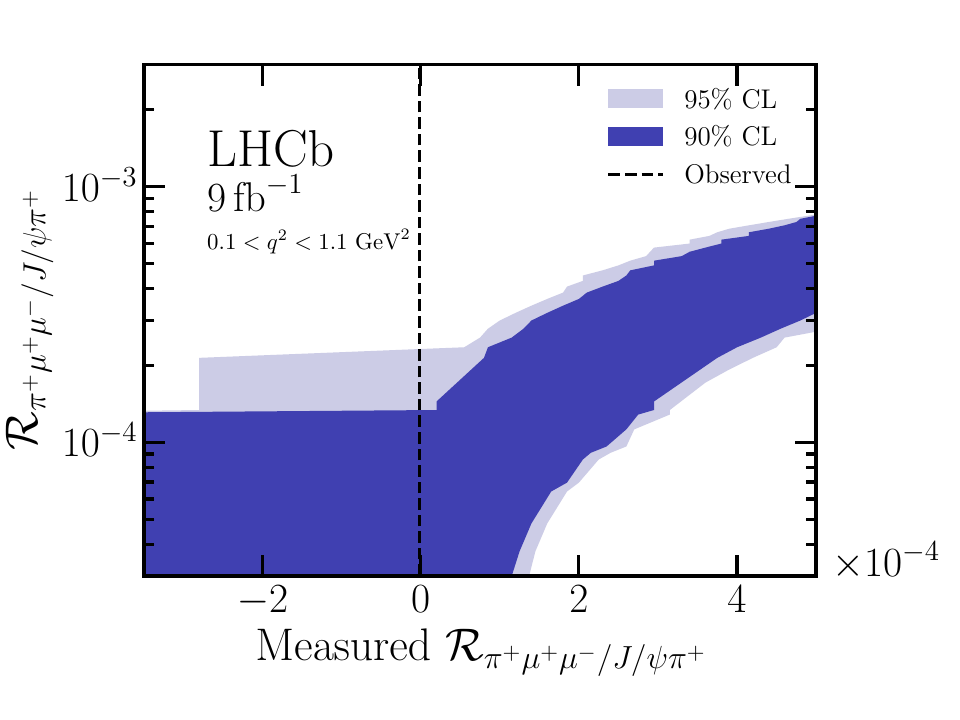} &  
    \includegraphics[width=0.49\linewidth]{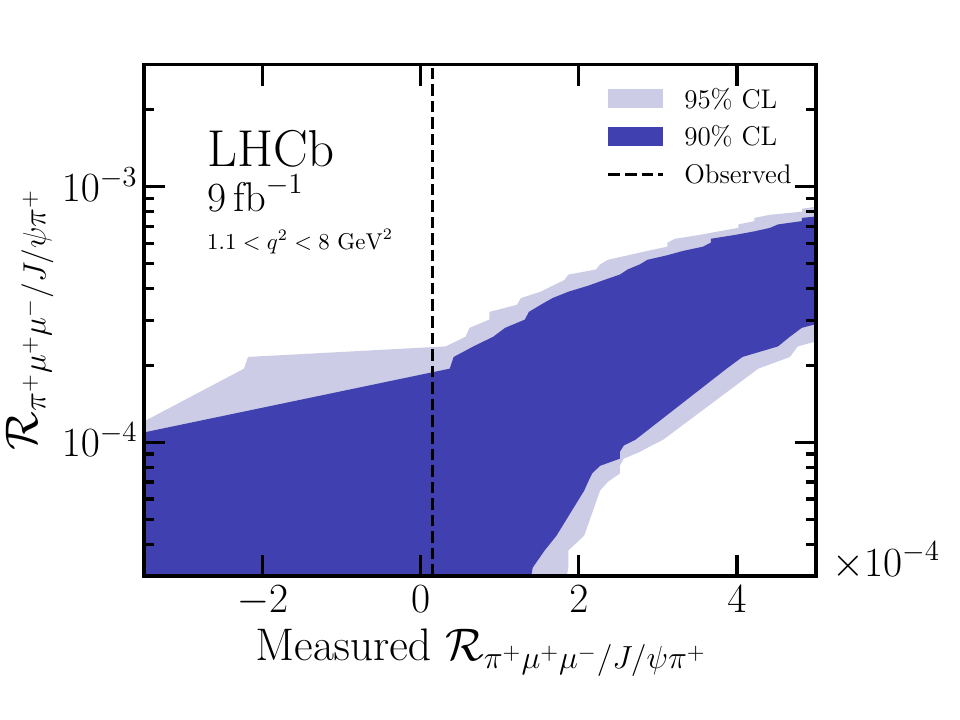} \\ 
    Low & Central \\
    
    \includegraphics[width=0.49\linewidth]{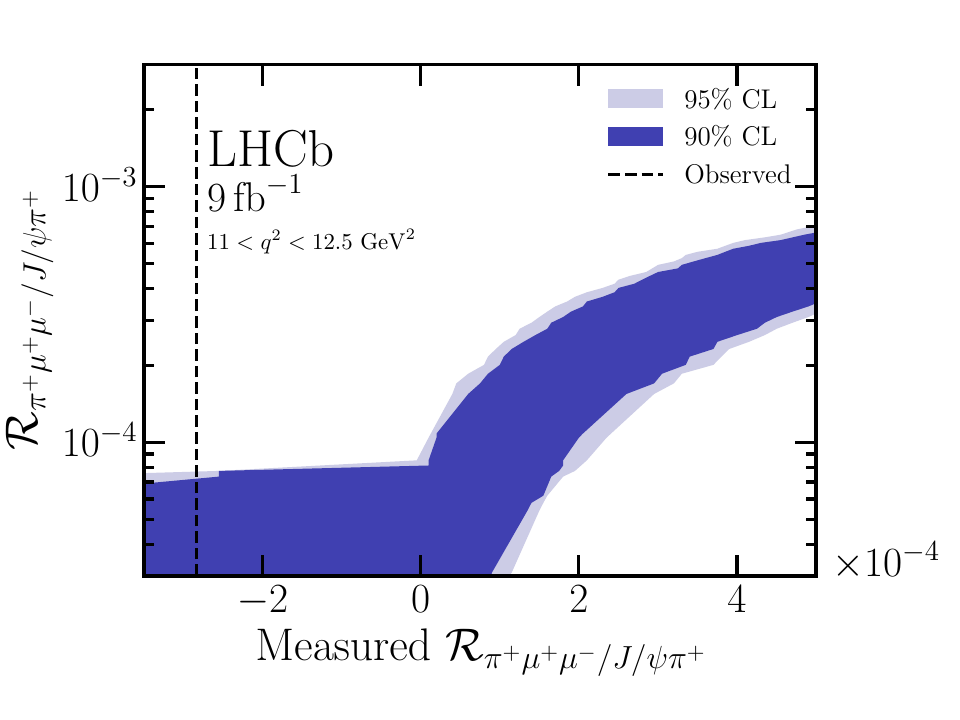} &
    \includegraphics[width=0.49\linewidth]{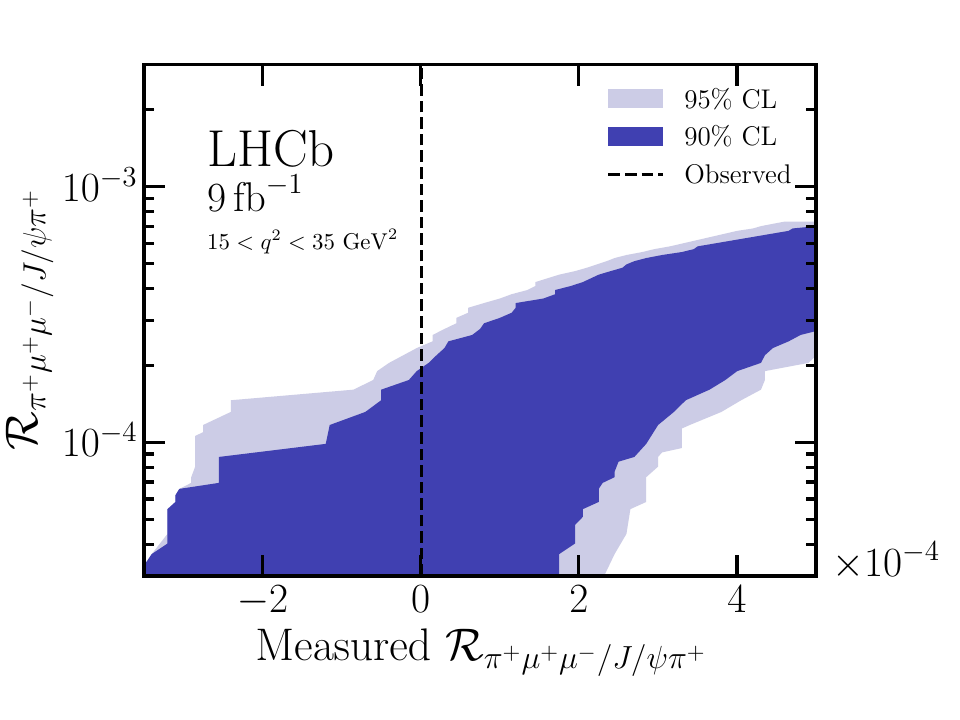} \\
    Intermediate & High \\
    
    \multicolumn{2}{c}{\includegraphics[width=0.49\linewidth]{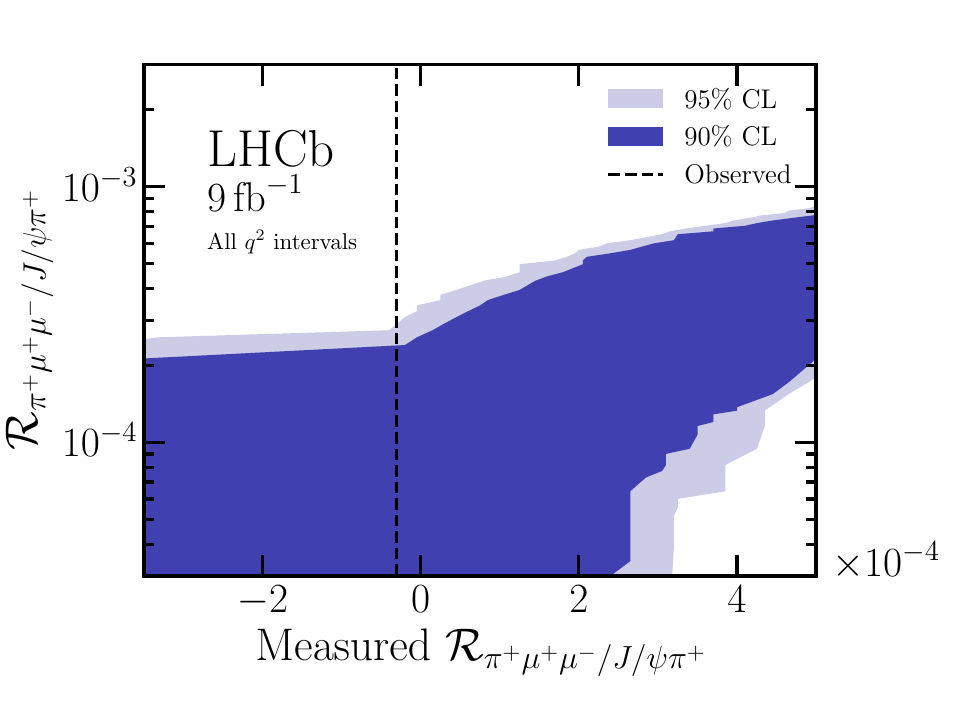}} \\
    \multicolumn{2}{c}{All}
    \end{tabular}
    \caption{
      Confidence belts generated using pseudoexperiments according to the Feldman--Cousins prescription~\cite{Feldman:1997qc} for each $q^2$ interval and for all intervals combined.
    The vertical dashed line shows the central value from the fit to data.}
    \label{fig:conf_bands}
\end{figure}

\begin{figure}
    \centering
    \includegraphics[width=0.7\textwidth]{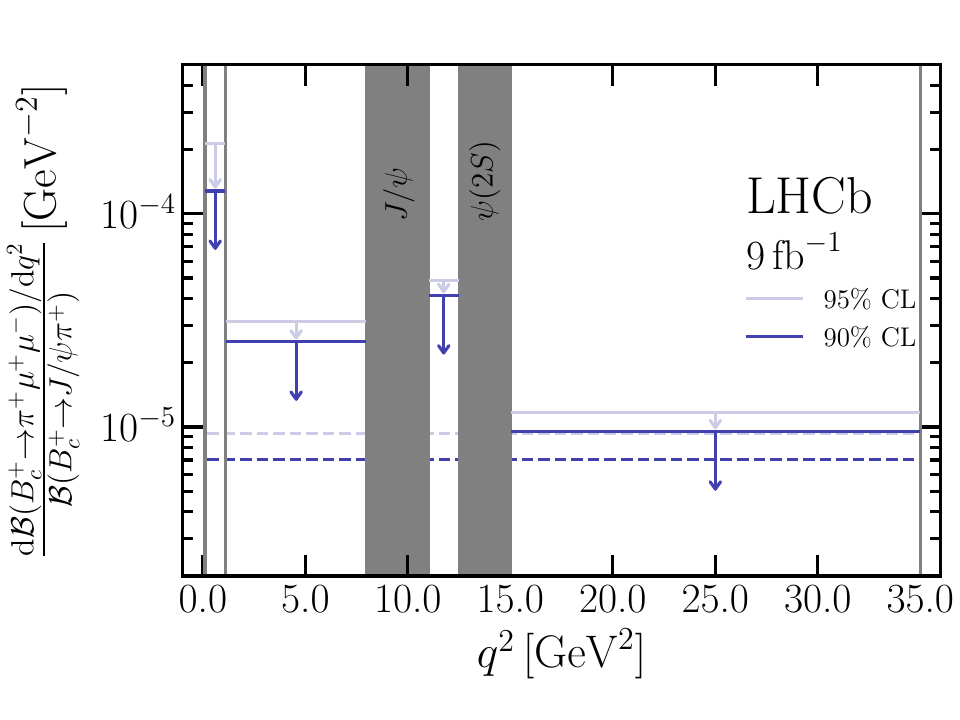}
    \caption{Upper limits on the normalised differential branching fraction for nonresonant \mbox{\decay{\Bc}{\pip\mumu}} decays as a function of $q^2$. 
    The solid lines show the results for each $q^2$ bin, while the dashed lines show the results for all bins combined.}
    \label{fig:summary_q2differential}
\end{figure}

Upper limits on the branching fraction ratios are obtained following the Feldmann--Cousins prescription~\cite{Feldman:1997qc}: pseudoexperiments are generated for various values of $R_{\pip\mumu/\jpsi\pip}$ and the resulting distribution of measured $R_{\pip\mumu/\jpsi\pip}$ is used to form confidence belts. 
Figure~\ref{fig:conf_bands} shows confidence belts at $90\%$ and $95\%$ confidence level (CL). 
Table~\ref{tab:pimumu_fit_to_data_unblind} gives the results for $R_{\pip\mumu/\jpsi\pip}$ and the obtained limits.
To assess the impact of the systematic uncertainties, the fits are repeated fixing the nuisance parameters to their central values. 
Figure~\ref{fig:summary_q2differential} summarises the obtained limits on the normalised differential branching fraction.
As further checks the procedure is repeated restricting the signal yield to positive values, or performing nonextended maximum-likelihood fits. 
No significant changes in the obtained upper limits are found.

{
\renewcommand{\arraystretch}{1.2}
\begin{table}[htb!]
    \centering
    \caption{Results for $R_{\pip\mumu/\jpsi\pip}$, where the first uncertainties are statistical and the second are systematic, together with upper limits (ULs) at $90\%$ and $95\%$~CL.}    

 \begin{tabular}{r@{$\,< q^2 <\,$}lr@{$\,\times\,$}lcc}
 \toprule
   \multicolumn{2}{c}{$q^2$ interval}   &  
   \multicolumn{2}{c}{$R_{\pip\mumu/\jpsi\pip}$}  &
   \multicolumn{1}{c}{ UL at $90\%$ CL}  &
   \multicolumn{1}{c}{  UL at $95\%$ CL}  \\
    \midrule
    0.1 & $\phantom{1}1.1\gev^2$    & $(-0.2\,^{+\phantom{1}4.8}_{-\phantom{1}3.5}\,^{+\,0.8}_{-\,0.7})$ & $10^{-5}$  &  1.3$\times 10^{-4}$   &  2.1$\times 10^{-4}$  \\ 
    1.1 & $\phantom{1}8.0\gev^2$    & $(1.5\,^{+\phantom{1}7.9}_{-\phantom{1}6.4}\,^{+\,2.3}_{-\,1.4})$  & $ 10^{-5}$ &  1.7$\times 10^{-4}$   &  2.2$\times 10^{-4}$  \\ 
   11.0 & $12.5\gev^2$   & $(-28.4\,^{+10.5}_{-16.1}\,^{+\,1.4}_{-\,1.7})$                    & $10^{-5}$  &  0.6$\times 10^{-4}$   &  0.7$\times 10^{-4}$  \\ 
   15.0 & $35.0\gev^2$   & $(0.2\,^{+11.5}_{-10.5}\,^{+\,3.4}_{-\,2.3})$                      & $10^{-5}$ &  1.9$\times 10^{-4}$   & 2.3$\times 10^{-4}$  \\ 
\multicolumn{2}{c}{All} & $(-3.0\,^{+15.0}_{-13.8}\,^{+\,2.4}_{-\,4.8})$                     & $10^{-5}$  &  2.1$\times 10^{-4}$   &  2.7$\times 10^{-4}$  \\
\bottomrule
 \end{tabular}
 \label{tab:pimumu_fit_to_data_unblind}
\end{table}
}

For the resonant modes, the yields obtained from the fits and the efficiency ratio are used to calculate the fraction
\begin{equation*}
    F_{\psitwos/\jpsi} = \frac{N_{\psitwos\pip}}{N_{\jpsi\pip}} \cdot \frac{\varepsilon_{\jpsi\pip}}{\varepsilon_{\psitwos\pip}}\,\text{,}
\end{equation*}
which is translated into the ratio of branching fractions 
\begin{equation*}
    R_{\psitwos/\jpsi} \equiv \frac{\BF(\decay{\Bc}{\psitwos\pip})}{\BF(\decay{\Bc}{\jpsi\pip})} = F_{\psitwos/\jpsi} \cdot \frac{\BF(\decay{\jpsi}{\mumu})}{\BF(\decay{\psitwos}{\mumu})} \,\text{.}
\end{equation*}
The branching fractions of the \decay{\jpsi}{\ep\en} and \decay{\psitwos}{\ep\en} decays are used in place of $\BF(\decay{\jpsi}{\mumu})$ and $\BF(\decay{\psitwos}{\mumu})$ because the branching fraction of electronic \psitwos decays is more precisely measured than its muonic counterpart and the decay rates into $\ep\en$ and \mumu states are expected to be identical, up to negligible mass-dependent corrections, due to the universality of the electroweak couplings to charged leptons.
The results are 
\begin{eqnarray*}
     F_{\psitwos/\jpsi}   &=& 0.0337\pm 0.0024\stat \pm 0.0004\syst\text{,}          \\
    R_{\psitwos/\jpsi}    &=& 0.254\pm 0.018 \stat \pm 0.003 \syst \pm 0.005\,({\rm BF})\,,
\end{eqnarray*}
where for $R_{\psitwos/\jpsi}$ the third uncertainty is due to limited precision of the known branching fractions for $\jpsi$ and $\psitwos$ leptonic decays~\cite{PDG2022}. 
The results are consistent with, but more precise than, those obtained in previous analyses~\cite{LHCb-PAPER-2012-054,LHCb-PAPER-2015-024}. 
As a cross check the values of $F_{\psitwos/\jpsi}$ and $R_{\psitwos/\jpsi}$ are determined at the $\pip\mumu$ WP, obtaining consistent results. 

\section{Summary} 
\label{sec:Summary} 

A search for nonresonant \decay{\Bc}{\pip\mumu} decays is performed together with an updated measurement of the ratio of the \mbox{\decay{\Bc}{\psitwos\pip}} and \decay{\Bc}{\jpsi\pip} branching fractions. 
The analysis uses proton-proton collision data collected with the LHCb detector between 2011 and 2018, corresponding to an integrated luminosity of 9\invfb.
No evidence for an excess of signal events over background is observed for nonresonant \decay{\Bc}{\pip\mumu} decays and an upper limit is set on the branching fraction ratio
\begin{equation*}
 \frac{\BF(\decay{\Bc}{\pip\mumu})}{\BF(\decay{\Bc}{\jpsi\pip})} < 2.1\times 10^{-4}\text{,} 
\end{equation*}
at $90\%$ confidence level.
This is the first limit on \Bc~decays mediated only by annihilation diagrams into a semileptonic final state.
For the resonant \decay{\Bc}{\psitwos\pip} mode, the branching fraction ratio is measured to be 
\begin{equation*}
    \frac{\BF(\decay{\Bc}{\psitwos\pip})}{\BF(\decay{\Bc}{\jpsi\pip})} = 0.254\pm 0.018\stat \pm 0.003\syst \pm 0.005\,({\rm BF})\text{,}
\end{equation*}
where the third uncertainty is due to limited precision of the known branching fractions for $\jpsi$ and $\psitwos$ leptonic decays~\cite{PDG2022}. 
This measurement is consistent with, and supersedes, previous LHCb results on the same quantity~\cite{LHCb-PAPER-2012-054,LHCb-PAPER-2015-024} and is the most precise to date.

\clearpage

\section*{Acknowledgements}
%
%
\noindent We express our gratitude to our colleagues in the CERN
accelerator departments for the excellent performance of the LHC. We
thank the technical and administrative staff at the LHCb
institutes.
We acknowledge support from CERN and from the national agencies:
CAPES, CNPq, FAPERJ and FINEP (Brazil); 
MOST and NSFC (China); 
CNRS/IN2P3 (France); 
BMBF, DFG and MPG (Germany); 
INFN (Italy); 
NWO (Netherlands); 
MNiSW and NCN (Poland); 
MCID/IFA (Romania); 
MICINN (Spain); 
SNSF and SER (Switzerland); 
NASU (Ukraine); 
STFC (United Kingdom); 
DOE NP and NSF (USA).
We acknowledge the computing resources that are provided by CERN, IN2P3
(France), KIT and DESY (Germany), INFN (Italy), SURF (Netherlands),
PIC (Spain), GridPP (United Kingdom), 
CSCS (Switzerland), IFIN-HH (Romania), CBPF (Brazil),
and Polish WLCG (Poland).
We are indebted to the communities behind the multiple open-source
software packages on which we depend.
Individual groups or members have received support from
ARC and ARDC (Australia);
Key Research Program of Frontier Sciences of CAS, CAS PIFI, CAS CCEPP, 
Fundamental Research Funds for the Central Universities, 
and Sci. \& Tech. Program of Guangzhou (China);
Minciencias (Colombia);
EPLANET, Marie Sk\l{}odowska-Curie Actions, ERC and NextGenerationEU (European Union);
A*MIDEX, ANR, IPhU and Labex P2IO, and R\'{e}gion Auvergne-Rh\^{o}ne-Alpes (France);
AvH Foundation (Germany);
ICSC (Italy); 
GVA, XuntaGal, GENCAT, Inditex, InTalent and Prog.~Atracci\'on Talento, CM (Spain);
SRC (Sweden);
the Leverhulme Trust, the Royal Society
 and UKRI (United Kingdom).

\addcontentsline{toc}{section}{References}
\bibliographystyle{LHCb}
\bibliography{main,standard,LHCb-PAPER,LHCb-CONF,LHCb-DP,LHCb-TDR}
 
\newpage
\centerline
{\large\bf LHCb collaboration}
\begin
{flushleft}
\small
R.~Aaij$^{35}$\lhcborcid{0000-0003-0533-1952},
A.S.W.~Abdelmotteleb$^{54}$\lhcborcid{0000-0001-7905-0542},
C.~Abellan~Beteta$^{48}$,
F.~Abudin{\'e}n$^{54}$\lhcborcid{0000-0002-6737-3528},
T.~Ackernley$^{58}$\lhcborcid{0000-0002-5951-3498},
B.~Adeva$^{44}$\lhcborcid{0000-0001-9756-3712},
M.~Adinolfi$^{52}$\lhcborcid{0000-0002-1326-1264},
P.~Adlarson$^{78}$\lhcborcid{0000-0001-6280-3851},
C.~Agapopoulou$^{46}$\lhcborcid{0000-0002-2368-0147},
C.A.~Aidala$^{79}$\lhcborcid{0000-0001-9540-4988}
% ...

{\footnotesize \it

% ...
$^{35}$Nikhef National Institute for Subatomic Physics, Amsterdam, Netherlands\\
% ...
$^{44}$Instituto Galego de F{\'\i}sica de Altas Enerx{\'\i}as (IGFAE), Universidade de Santiago de Compostela, Santiago de Compostela, Spain\\
% ...
$^{46}$European Organization for Nuclear Research (CERN), Geneva, Switzerland\\
% ...
$^{48}$Physik-Institut, Universit{\"a}t Z{\"u}rich, Z{\"u}rich, Switzerland\\
% ...
$^{52}$H.H. Wills Physics Laboratory, University of Bristol, Bristol, United Kingdom\\
% ...
$^{54}$Department of Physics, University of Warwick, Coventry, United Kingdom\\
% ...
$^{57}$School of Physics and Astronomy, University of Glasgow, Glasgow, United Kingdom\\
$^{58}$Oliver Lodge Laboratory, University of Liverpool, Liverpool, United Kingdom\\
% ...
$^{66}$Syracuse University, Syracuse, NY, United States\\
% ...
$^{78}$Department of Physics and Astronomy, Uppsala University, Uppsala, Sweden, associated to $^{57}$\\
$^{79}$University of Michigan, Ann Arbor, MI, United States, associated to $^{66}$\\
% ...
}
\end{flushleft}

\end{document}